\documentclass[a4paper, 10pt, aps, showkeys, showpacs, nofootinbib, twocolumn]{revtex4-2}
\usepackage{graphicx}
\usepackage{amsfonts}
\usepackage{amssymb}
\usepackage{amsbsy}
\usepackage{amsmath}
\usepackage{mathrsfs}
\usepackage{latexsym}
\usepackage{bm}
\usepackage{subfigure} 
\usepackage{wasysym}
\usepackage{mathbbol}
\usepackage{bigints} 
\usepackage{fontawesome}
\usepackage{natbib}
\usepackage{xcolor}
\usepackage{footnote}
\usepackage[colorlinks = true, linkcolor=blue, citecolor=cyan, urlcolor=blue]{hyperref}
\usepackage{orcidlink}
\begin{document}
%%%%%%%%%%%%%%%%%%%%%%%%%%%%%%%%%%%%%%%%%%%%%%%%%%%%%%%%%%%

\title[Accretion flow around black holes]{General relativistic viscous accretion flow around Konoplya-Zhidenko black hole}

\author{Subhankar Patra \orcidlink{0000-0001-7603-3923}}
\email{psubhankar@iitg.ac.in}

\author{Bibhas Ranjan Majhi \orcidlink{0000-0001-8621-1324}}
\email{bibhas.majhi@iitg.ac.in}

\author{Santabrata Das \orcidlink{0000-0003-4399-5047}}
\email{sbdas@iitg.ac.in}

\affiliation{Department of Physics, Indian Institute of Technology Guwahati, Guwahati 781039, Assam, India}

\date{\today}

% Abstract of the paper
\begin{abstract}
	We investigate the properties of accretion flows around the Konoplya-Zhidenko (KZ) black hole, which is proposed by deforming the Kerr metric with a single deformation parameter to test the no-hair theorem using gravitational wave observations. The dynamical equations describing the general relativistic viscous accreting flow are solved self-consistently to find the transonic accretion solutions in terms of global constants, such as energy ($E$), angular momentum ($\mathcal{L}$), viscosity parameter ($\alpha$), spin ($a_{k}$), and deformation parameter ($\eta_0$). We obtain five distinct types of accretion solutions (O, A, $\text{A}^{\prime}$, W, and I-types), and observe that those solutions are not unique but rather continue to exist for wide range of parameter spaces in the $\mathcal{L}-E$ plane. Furthermore, we find that the viscous accretion flows can harbor shock waves when the relativistic shock conditions are satisfied. Consequently, the shock-induced global accretion solutions are obtained, and the effect of $\eta_0$ on shock properties, such as shock radius ($r_{\rm sh}$) and change in electron temperature ($T_{\rm e}$) across the shock front are investigated. Moreover, we calculate the spectral energy distributions (SEDs) of accretion flow using the relativistic thermal bremsstrahlung emission coefficient and study the modification of SEDs due to the increase of $\eta_0$ for both shock-induced and shock-free solutions. In addition, it has been noticed that the observable quantities, like quasi-periodic oscillation frequency ($\nu_{\rm QPO}$) and bolometric disc luminosity ($L$), are strongly dependent on $\eta_0$. Finally, we phenomenologically show that the KZ black hole is consistent with the high-frequency QPOs, commonly observed in black hole binaries and black hole candidates.
\end{abstract}

%Select between one and six entries from the list of approved keywords. 


%\pacs{42.50.Ex, 32.80.Wr, 32.80.-t; 32.10.Fn}

\maketitle
%%%%%%%%%%%%%%%%%%%%%%%%%%%%%%%%%%%%%%%%%%%%%%%%%%
  
\section{Introduction}
In the realm of Einstein’s general relativity (GR), we face some well-known issues, like the nature of singularities, dark energy, dark matter, quantization of gravitational interactions, etc. To address these, theorists are engaged in developing the alternative theories of gravity, such as modified gravity theory \cite{Sullivan-2019}, higher curvature \cite{Deser-2002}, brane world gravity \cite{Maartens-2010}, etc. Also, several parametric deviations from the Kerr metric have been proposed in the literature. Such non-Kerr metrics are described by few additional parameters along with the mass and spin of the black holes \cite{Nampalliwar-2020}. Moreover, they are not the solutions of Einstein's field equations; otherwise it would contradict the validity of no-hair theorem \cite{Carter-1971, Robinson-1975}. They are classified into two categories: top-down metrics and bottom-up metrics. The top-down metrics are obtained as the solutions of alternative gravity theories, such as Kerr-Sen black holes \cite{Sen-1992}, Einstein-dilaton-Gauss-Bonnet (EDGB) black holes \cite{Kanti-1995}, and Chern-Simons black holes \cite{Yunes-2009-084043}. In contrast, bottom-up metrics are not derived from a specific gravity theory; instead, they are formulated just by incorporating the parametric deviation functions into the usual Kerr metric \cite{johannsen-2010, johannsen-2011, vigeland-2011, Johannsen-2013, Rezzolla-2014, konoplya-2016b, konoplya_2016, Ma-2024}. Note that the underlying gravity theories for the bottom-up metrics are still unknown. In this work, we focus on some bottom-up metrics.

In recent years, numerous studies have focused on the observational testings of the Kerr hypothesis using bottom-up metrics, aiming to put constraints on the deformation parameters \cite{Ni-2016, konoplya-2016b,Cardenas-Avendano-2019, EHT-2020, Nampalliwar-2020, Shashank-2021, Bambi-2023, Tripathi-2024}. Indeed, these studies imply that such non-Kerr black holes are consistent with the modern-day observations of both electromagnetic and gravitational-wave signals. Therefore, the issue of whether such an alternative prescription is required remains unsettled, mainly due to the absence of precise observational data. Moreover, it will be helpful from the observational point of view, if the theoretical predictions corresponding to the individual astrophysical phenomenon can be listed. Here, we choose accretion flows onto the black holes which is the most acceptable physical process for energy extraction in astrophysical sources like active galactic nuclei (AGN) and black hole X-ray binaries (BH-XRBs) \cite{Pringle-1981, Frank-2002}. To understand their observed spectrum features, a vast number of theoretical models have been proposed on the topic of accretion flow based on different physical conditions in the strong gravity regime \cite[and references therein]{netzer-2013, abramowicz-2013, yuan-2014}. Indeed, the spectral analysis helps to determine the black hole parameters ($i.e$., mass and spin) \cite{nandi-2018, das-2021, mondal-2022a, mondal-2022b, palit-2023, mondal-2023, heiland-2023, mondal-2024} and some fundamental properties of the accretion disc as well (e.g., mass accretion and outflow rates, Comtonizing corona size, quasi-periodic oscillation (QPO) frequency, disc inclination angle) \cite{nandi-2018, sreehari-2020, das-2021, Sriram-2021, majumder-2022, mondal-2022a, mondal-2022b, palit-2023, mondal-2023, heiland-2023, dhaka-2023, Rawat-2023a, Rawat-2023c, mondal-2024}. In this venture, we have already studied the accretion flow in the Johannsen-Psaltis (JP) type deformed spacetime \cite{Patra-2022, Patra-2023}, a bottom-up case. Recently, another parametric deviation to the Kerr metric is introduced by Konoplya and Zhidenko to test the GR in strong gravity regime using the observations of gravitational waves (GWs) resulting from the binary black hole mergers \cite{konoplya_2016}, and we hereafter refer to it as the KZ metric, which is also a bottom-up case. It is essential to mention that Konoplya, Rezzolla, and Zhidenko (KRZ) construct a new parameterized framework to obtain a generic stationary and axisymmetric black hole spacetime \cite{konoplya-2016b}. They validated their approach by accurately reproducing several known rotating black holes, such as Kerr, Kerr-Sen, and EDGB black holes. However, a notable open question remains: can the KZ black hole be reproduced using the KRZ parametrization framework? We do not address this question in this work because our focus is only on the accretion dynamics. However, we encourage readers to explore the KRZ parametrization framework for reproducing the KZ black holes. Towards this, for the first time, to the best of our knowledge, we study the physical properties of accretion disc in the KZ spacetime by considering hydrodynamical aspects of accretion flow. We compare the obtained results with our earlier findings (see, \cite{Patra-2022, Patra-2023}) for the JP black holes, which makes a qualitative distinction between the type of deformations for these two metrics.

The KZ metric is characterized by the mass, spin, and a deformation parameter ($\eta_0$). This metric reverts back to the original Kerr metric for $\eta_0 \rightarrow 0$ limit. We briefly mention the motivation behind the formation of KZ spacetime. After the detection of gravitational waves by LIGO and VIRGO collaborations~\cite{abbott-2016a, abbott-2016-b}, a very good agreement is established between the observed waveform of the gravitational-wave signal and the results obtained from the numerical simulation of the inspiral and merger of two black holes in Einstein's gravity. However, due to inaccuracy in detecting the gravitational-wave profile, a wide range of mass and spin parameters of the black hole in numerical simulation provide the same gravitational-wave signal~\cite{abbott-2016a, abbott-2016-b}. This situation can also be replicated by an intuitive idea of introducing a parametric deviation from the Kerr spacetime instead of changing the given black hole parameters in the same Kerr spacetime. In \cite{konoplya_2016}, authors validated the above mentioned intuitive notion, illustrating that the inclusion of a non-negligible deformation (through $\eta_0$) into the original Kerr metric yields a quasinormal frequency that is closely matched with the result obtained from the Kerr geometry. Therefore, both the Kerr and KZ non-Kerr spacetime are consistent with the ringdown phase of gravitational waves. As the ringdown picture is in the regime of strong gravitational field, we can not rule out the possibility of formation of a non-Kerr black hole. The only way to do that is by improving the precision level of detecting the gravitational-wave profile, which will furnish the spacetime parameters of the final black hole with high accuracy.

 It is worth noting that while the deformation in KZ metric does not alter the asymptotic properties of the spacetime, it does impact the near-horizon geometry, notably shifting the location of event horizon. Recently, interest has grown in examining the influence of KZ deformation on different astrophysical phenomena \cite{Barman-2021}. However, in this work, our concern is to explore the accretion flow around the KZ black holes. As the deformation primarily alters the near-horizon geometry, it will also impact the dynamics of accretion flows, and thereby change the accretion disc properties. With this objective in mind, we investigate the effect of $\eta_0$ on the properties of viscous accretion flow. We utilize the framework of general relativistic hydrodynamics to study the dynamics of accretion flow \cite{Rezzolla-2013}. We start our analysis by finding distinct classes of global transonic accretion solutions in terms of flow energy ($E$) and angular momentum ($\mathcal{L}$), and accordingly divide the $\mathcal{L}-E$ parameter space. We then find the shock-induced accretion solutions using the relativistic shock conditions \cite{Taub-1948}. Moreover, we calculate the spectral energy distribution (SED) corresponding to the accretion solutions by using the emissivity of relativistic thermal bremsstrahlung radiation \cite{novikov-1973}. Further, we examine how the shock location ($r_{\rm sh}$), QPO oscillation frequency ($\nu_{\rm QPO}$), and bolometric disc luminosity ($L$) change with $\eta_0$. Since the modulation of the shock front yields the QPO phenomena, we phenomenologically find the parameter spaces in $\mathcal{L}-\eta_0$ plane that admit observed high-frequency QPOs (HFQPOs) for the black hole binary and black hole candidate systems. Finally, we distinguish the KZ and JP deformations by comparing their behavior in effecting the accretion disc properties.

We arrange the paper as follows. In Section \ref{sec:model_equations}, we formulate the mathematical background that governs the accretion disc in the KZ spacetime. In Section \ref{sec:result}, we present the transonic accretion solutions with and without shocks, and also discuss the effect of deformation on various shock properties. In Section \ref{sec:HFQPO}, we examine a phenomenology on the observed HFQPOs for black hole binaries and black hole candidates. Finally, in Section \ref{sec:conclusion}, we present the concluding remarks.

\section{Model equations governing accretion disc} 
\label{sec:model_equations}
We explore the accretion dynamics of a viscous flow, including shock transition in a stationary and axisymmetric KZ black hole spacetime. The required model equations are developed under the framework of general relativistic hydrodynamics \cite{Rezzolla-2013}. Here, we follow the analysis of \cite{Dihingia-2019}; however, the current equations contain an additional parameter $\eta_0$ that we need to explicitly mention.

The line element of Konoplya-Zhidenko (KZ) non-Kerr spacetime can be expressed in Boyer-Lindquist coordinates ($t, r, \theta, \phi$) as \cite{konoplya_2016},
\begin{equation}
	\begin{split}
		\label{eq:deformed-Kerr-metric}
		ds^{2} = &-\frac{N^{2}(r, \theta) - W^{2}(r, \theta)\sin^{2}\theta}{K^{2}(r, \theta)}dt^{2}
		\\& - 2W(r, \theta)r\sin^{2}\theta dtd\phi + K^{2}(r, \theta)r^{2}\sin^{2}\theta d\phi^{2}
		\\& + \frac{\Sigma(r, \theta)}{r^{2}N^{2}(r, \theta)}dr^{2} + \Sigma(r, \theta)d\theta^{2},
	\end{split}
\end{equation}
where 
\begin{equation}
	\begin{split}
		& N^2(r, \theta) = \frac{\Delta}{r^2} - \frac{\eta_0}{r^3}, \Delta = r^{2} - 2M_{\rm BH}r + a_{\rm k}^{2},\\
		& W(r, \theta) = \frac{2a_{\rm k}M_{\rm BH}}{\Sigma} + \frac{\eta_0 a_{\rm k}}{r^2\Sigma}, \Sigma = r^2 + a_{\rm k}^2\cos^{2}\theta,\\
		& K^2(r, \theta) = \frac{(r^{2} + a_{\rm k}^{2})^{2} - \Delta a_{\rm k}^{2} \sin^{2}\theta}{r^{2}\Sigma} + \frac{a_{\rm k}^{2}\eta_0\sin^{2}\theta}{r^{3}\Sigma}.
	\end{split}
\end{equation} 
In the above expressions, $M_{\rm BH}$ and $a_{\rm k}$ denote the mass and spin of the black hole (BH), respectively, and $\eta_0$ is the deformation parameter. For $\eta_0 = 0$, the metric in Eq.~(\ref{eq:deformed-Kerr-metric}) reduces to the original Kerr metric~\cite{Boyer_1966}. In this work, we use a unit system $G = M_{\rm BH} = c = 1$, where $G$ is the gravitational constant, and $c$ is the speed of light. We use a condition $g^{rr}= 0$ to find the event horizon location ($r_{\rm H}$), which can be calculated numerically by finding the maximum root of the equation $r_{\rm H}^{3} - 2r_{\rm H}^{2} + a_{\rm k}^{2}r_{\rm H} - \eta_0 = 0$. Here, we restrict our analysis within the allowed range of $a_{\rm k}$ and $\eta_0$ as delineated in \cite{konoplya_2016}.

For viscous fluid, the energy-momentum tensor is given by \cite{landau-2013},
\begin{equation}
	\label{eq:energy-momentum-tensor}
	T^{ik} = (e + p)u^{i}u^{k} + pg^{ik} + \pi^{ik},
\end{equation}
where $e$ ($= \rho + \Pi$, $\rho$ is the mass density, and $\Pi$ is the internal energy density) is the total energy density, $p$ is the isotropic pressure, $u^{i}$ is the velocity 4-vector, $g^{ik}$ is the inverse metric tensor, $\pi^{ik} = - 2\eta\sigma^{ik}$ is the viscous stress tensor, and $\eta$ is the dynamic viscosity. The shear tensor $\sigma^{ik}$ can be written as~\cite{Peitz-1996},
\begin{equation}
	\label{eq:shear-tensor}
	\begin{split}
	\sigma^{ik} = \frac{1}{2}\left[(\nabla^{i}u^{k} + \nabla^{k}u^{i} + a^{i}u^{k} + a^{k}u^{i}) - \frac{2}{3}\Theta_{\rm exp}\mathfrak{h}^{ik} \right],
	\end{split}
\end{equation}
where $a^{i} = u^{l}\nabla_{l}u^{i}$ is the acceleration 4-vector, $\Theta_{\rm exp}$ ($= \nabla_{l}u^{l}$) is the expansion of fluid world line, and $\mathfrak{h}^{ik}$ ($= u^{i}u^{k} + g^{ik}$) is the projection tensor.

The conservation equation for the energy-momentum tensor is given by $\nabla_{k}T^{ik} = 0$. When we project this equation onto the projection tensor $\mathfrak{h}^{l}_{i}$, we obtain the momentum equation (where $l$ only takes spatial coordinates) as,
\begin{equation}
	\label{eq:momentum-equation}
	(e + p)u^{k}\nabla_{k}u^{l} + (g^{l k} + u^{l}u^{k})\nabla_{k}p + \mathfrak{h}^{l}_{i}\nabla_{k}\pi^{ik} = 0.
\end{equation}
However, when we project $\nabla_{k}T^{ik} = 0$ onto $u_{i}$, which is perpendicular to $\mathfrak{h}^{i}_{l}$ (i.e., $\mathfrak{h}^{i}_{l}u_i = 0$), we derive the energy equation. Subsequently, by incorporating the mass flux conservation equation $\nabla_{k}(\rho u^{k}) = 0$ into the energy equation, we can express it in the form,
\begin{equation}
	\label{eq:energy-equation}
	u^{k}\left(h_0\nabla_{k}\rho - \nabla_{k}e\right) = Q^{+},
\end{equation}
where $h_0$ ($= (e + p)/\rho$) is the specific enthalpy, and $Q^{+}$ ($= - u_{i}\nabla_{k}\pi^{ik} = \pi^{ik}\sigma_{ik}$) is the viscous heating term. In order to emphasize the impact of viscous dissipation, we exclude the radiative cooling from the above mentioned energy equation \cite{Chattopadhyay_2016}.

We consider the fluid motion is confined in the equatorial plane ($i.e, \theta = \pi/2$) of the black hole. Also, we assume that the flows obey same symmetries as that of the spacetime. Under these assumptions, the present Killing symmetries of the spacetime in Eq.~(\ref{eq:deformed-Kerr-metric}) give two conserved quantities along the flow direction as,
\begin{align}
	\label{eq:L}
	& \mathcal{L} = h_0u_{\phi} - \frac{2\nu\sigma^{r}_{\phi}}{u^{r}},\\
	\label{eq:E}
	& E = -\left(h_0u_{t} - \frac{2\nu\sigma^{r}_{t}}{u^{r}}\right),
\end{align}	
where $\mathcal{L}$ is the bulk angular momentum per unit mass, $E$ is the Bernoulli constant, and $\nu$ ($= \eta/\rho = \alpha C_{s}H$) \cite{Gammie_1997, Popham_1998} is the kinematic viscosity. Here, $\alpha$ is the viscosity parameter, $C_{s}$ ($= \sqrt{\Gamma p/(e + p)}$) is the sound speed with adiabatic index $\Gamma$, and $H$ ($= \sqrt{pr^{3}/(\rho F)}$) \cite{lasota-1994, riffert-1995, Peitz-1996} is the half thickness of disc. The quantity $F$ is given by in terms of flow angular velocity $\Omega$ ($= u^{\phi}/u^{t}$) and specific angular momentum $\lambda$ ($ = - u_{\phi}/u_{t}$) as $F = F_1/(1 - \Omega \lambda)$, where $F_1 = ((r^{2} + a_{\rm k}^2)^{2} + 2\Delta a_{\rm k}^{2})/((r^{2} + a_{\rm k}^2)^{2} - 2\Delta a_{\rm k}^{2})$. We evaluate the shear tensor components $\sigma^{r}_{\phi}$ and $\sigma^{r}_{t}$ using Eq.~(\ref{eq:shear-tensor}) as,
\begin{align}
	\label{eq:stress-tensor-r-phi-component}
	\sigma^{r}_{\phi} = \frac{1}{2}\left(A_{1} + A_{2}\frac{dv}{dr} + A_{3}\frac{d\lambda}{dr}\right),\\
	\label{eq:stress-tensor-r-t-component}
	\sigma^{r}_{t} = \frac{1}{2}\left(B_{1} + B_{2}\frac{dv}{dr} + B_{3}\frac{d\lambda}{dr}\right),
\end{align}	
where the quantities $A_{1}$, $A_{2}$, $A_{3}$, $B_{1}$, $B_{2}$, and $B_{3}$ are the function of flow variables, and their explicit forms have been given in Appendix~\ref{appendix-A}. In Eqs.~(\ref{eq:stress-tensor-r-phi-component}) and~(\ref{eq:stress-tensor-r-t-component}), $v$ denotes the r-component of 3-velocity in a co-rotating frame (which rotates with the flow angular velocity $\Omega$), and is obtained as $v = \sqrt{u^{r}u_{r}/(u^{t}u_{t}(\Omega \lambda - 1))}$ \cite{lu-1985}. Note that $v$ is negative for accretion.

The flow equations are manifested as second-order differential equations because of the first-order derivatives of $v$ and $\lambda$ in the expressions of $\sigma^{r}_{\phi}$ and $\sigma^{r}_{t}$. However, the second-order derivatives are negligible in the context of accretion flow dynamics \cite{Dihingia-2019}. Therefore, to simplify the numerical computation, we neglect the terms that possess first-order derivatives of the flow variables in Eqs.~(\ref{eq:stress-tensor-r-phi-component}) and (\ref{eq:stress-tensor-r-t-component}), which leads the flow equations as the first-order differential equations. Under this assumption, conservations of $\mathcal{L}$ and $E$ yield the following equations as,
\begin{align}
	\label{eq:L-conservation}
	&\frac{d\mathcal{L}}{dr} = \mathcal{L}_{0} + \mathcal{L}_{1}\frac{dv}{dr} + \mathcal{L}_{2}\frac{d\Theta}{dr} + \mathcal{L}_{3}\frac{d\lambda}{dr} = 0,\\ 
	\label{eq:E-conservation}
	&\frac{dE}{dr} = E_{0} + E_{1}\frac{dv}{dr} + E_{2}\frac{d\Theta}{dr} + E_{3}\frac{d\lambda}{dr} = 0,
\end{align}
where the explicit forms of the quantities $\mathcal{L}_{0}$, $\mathcal{L}_{1}$, $\mathcal{L}_{2}$, $\mathcal{L}_{3}$, $E_{0}$, $E_{1}$, $E_{2}$, and $E_{3}$ in terms of the flow variables are given in Appendix~\ref{appendix-B}. In Eqs.~(\ref{eq:L-conservation}) and (\ref{eq:E-conservation}), $\Theta = k_{\rm B}T/(m_{e}c^{2})$ denotes the dimensionless temperature, where $T$ is the flow temperature in Kelvin, $k_{\rm B}$ is the Boltzmann constant, and $m_{e}$ is the rest mass of electron. We calculate the gradients of $\Theta$ and $\lambda$ using Eqs.~(\ref{eq:L-conservation}) and (\ref{eq:E-conservation}) as,
\begin{align}
	\label{eq:Temperature-gradient}
	&\frac{d\Theta}{dr} = \Theta_{11}\frac{dv}{dr} + \Theta_{12},\\
	\label{eq:angular-momentum-gradient}
	&\frac{d\lambda}{dr} = \lambda_{11}\frac{dv}{dr} + \lambda_{12},
\end{align}
where
\begin{equation}
	\begin{split}
		& \Theta_{11} = \frac{L_{1}E_{3} - L_{3}E_{1}}{L_{3}E_{2} - L_{2}E_{3}}, \Theta_{12} = \frac{L_{0}E_{3} - L_{3}E_{0}}{L_{3}E_{2} - L_{2}E_{3}},\\
		& \lambda_{11} = \frac{L_{2}E_{1} - L_{1}E_{2}}{L_{3}E_{2} - L_{2}E_{3}}, \lambda_{12} = \frac{L_{2}E_{0} - L_{0}E_{2}}{L_{3}E_{2} - L_{2}E_{3}}.
	\end{split}
\end{equation}

As the gravitational acceleration is very high compared to the viscous acceleration along radial direction, we neglect the viscous acceleration term in the radial-momentum equation \cite{Gammie_1997, Popham_1998}. With this assumption, the radial-momentum equation is obtained by replacing $l = r$ in Eq.~(\ref{eq:momentum-equation}) as \cite{Dihingia-2019},
\begin{equation}
	\label{eq:radial-momentum}
	\gamma_{v}^{2}v\frac{dv}{dr} + \frac{1}{h_0\rho}\frac{dp}{dr} + \left(\frac{d\Phi^{\rm eff}}{dr}\right)_{\lambda} = 0,
\end{equation}
where $\gamma_{v} = 1/\sqrt{1 - v^2}$ is the Lorentz-factor for the radial 3-velocity, and $\Phi^{\rm eff} = - \frac{1}{2}\ln(\lambda f_{2} - f_{1})$ is the effective potential of the system. The quantities $f_{1}$ and $f_{2}$ are given by,
\begin{equation}
	\begin{split}
		& f_1 = - \frac{r^2(r^3 + a_{\rm k}^{2}(r+2)) - a_{\rm k} \lambda(2r^2 + \eta_0) + a_{\rm k}^{2}\eta_0}{r^{2}(\Delta r - \eta_0)},\\
		&f_2 = -\frac{a_{\rm k}(2r^2 + \eta_0) + \lambda (r^3 - 2r^2 - \eta_0)}{r^{2}(\Delta r - \eta_0)}.
	\end{split}
\end{equation}

Integrating the mass flux conservation equation $\nabla_{k}(\rho u^{k}) = 0$, we get the mass accretion rate as,
\begin{equation}
	\label{eq:mass_accretion}
	\dot{M} = - 4\pi v \gamma_{v}  \rho H\sqrt{(\Delta r - \eta_0)/r}.
\end{equation}
In a steady state, $\dot{M}$ remains constant throughout the disc. Therefore, using $\frac{d \dot{M}}{dr} = 0$, we rewrite Eq.~(\ref{eq:radial-momentum}) into the following form as,
\begin{equation}
	\label{eq:radial-momentum-1}
	R_{0} + R_{1}\frac{dv}{dr} + R_{2}\frac{d\Theta}{dr} + R_{3}\frac{d\lambda}{dr} = 0,
\end{equation} 
where $R_{0}$, $R_{1}$, $R_{2}$, and $R_{3}$ are given in Appendix~\ref{appendix-B}.

Thereafter, solving Eqs.~(\ref{eq:L-conservation}), (\ref{eq:E-conservation}) and (\ref{eq:radial-momentum-1}), we obtain the radial-velocity gradient as,
\begin{equation}
	\label{eq:velocity-gradient}
	\frac{dv}{dr} =  \frac{\mathcal N}{\mathcal D},
\end{equation}
where
\begin{equation}
	\begin{split}
	\mathcal{N} & = L_0(E_3R_2 - E_2R_3) + L_2(E_0R_3 - E_3R_0) \\
	& + L_3(E_2R_0 - E_0R_2),\\
	\mathcal{D} & = L_1(E_2R_3 - E_3R_2) + L_2(E_3R_1 - E_1R_3) \\ 
	& + L_3(E_1R_2 - E_2R_1).
	\end{split}
\end{equation}

Since $\pi^{ik}$ is negligible close to event horizon \cite{Peitz-1996}, we calculate the energy equation by neglecting the viscous heating term $Q^{+}$ in Eq.~(\ref{eq:energy-equation}) as,
\begin{equation}
	\label{eq:energy-equation-1}
	h_0\frac{d\rho}{dr} - \frac{de}{dr} = 0.
\end{equation}

For relativistic flows of variable $\Gamma$, we use an equation of state (EoS) with equal number densities of electron and ion as \cite{chattopadhyay-2009},
\begin{equation}
	\label{eq:EoS}
	\begin{split}
	e =\frac{\rho f}{1 + m_{p}/m_{e}},
	\end{split}
\end{equation}
where $m_{p}~(m_{e})$ is the rest mass of a proton (electron). Here, the quantity $f$ is expressed in terms of $\Theta$ as, 
\begin{equation}
	f = 1 + \frac{m_p}{m_e} + \Theta\left[\left(\frac{9\Theta + 3}{3\Theta + 2}\right) + \left(\frac{9\Theta + 3m_{p}/m_{e}}{3\Theta + 2m_{p}/m_{e}}\right)\right].
\end{equation}
Using this EoS, we get the entropy accretion rate of the flow very near the horizon after integrating Eq.~(\ref{eq:energy-equation-1}) as,
\begin{equation}
	\label{eq:entropy-accretion-rate}
	\begin{split}
		\mathcal{\dot{M}} & = - \exp{(\chi)} \Theta^{3/2}(3\Theta + 2)^{3/4}(3\Theta+2m_{p}/m_{e})^{3/4}\\ 
		& \times v\gamma_{v}H\sqrt{(\Delta r - \eta_0)/r},
	\end{split}
\end{equation}
where $\chi = (f - 1 - m_{p}/m_{e})/(2\Theta)$.
 
 In this work, we consider an isotopic emission of thermal bremsstrahlung radiation from the accretion disc. As the plasma temperature of the disc can reach $10^{12}~\rm K$ for hot accretion flow \cite{yuan-2014, dihingia-2020, das-2021, Sarkar-2022, Patra-2022}, electron-electron emission surpasses electron-ion emission \cite{Svensson-1982, nozawa-2009, Yarza-2020, Patra-2023}. So, we utilize the relativistic emission coefficient, which includes both the relativistic effect and the electron-electron emission in addition to the electron-ion emission. An approximate expression of the relativistic thermal bremsstrahlung emissivity at a particular emission frequency $\nu_e$ is given by~\cite{novikov-1973},
 \begin{equation}
 	\begin{split}
 		\mathcal{E}_{\nu_e}^{\rm ff} & =~6.8 \times 10^{-38} (\rho/m_p)^2Z^{2}T_{e}^{-1/2}(1 + 4.4\times 10^{-10}T_{e})\\& \times e^{-h\nu_e/k_{\rm B}T_{e}}\bar{g}_{\rm B}~{\rm erg~s^{-1}~cm^{-3}~Hz^{-1}},
 	\end{split}
 	\label{eq:emissivity}
 \end{equation}
 where $Z$ is the atomic number of ion (which is $1$ for hydrogen plasma), $T_{e}$ is the electron temperature (which relates with the flow temperature $T$ as $T_{e} = \sqrt{m_e/m_p}T$~\cite{chattopadhyay-2000}), and $h$ is the Planck constant. Here, $\bar{g}_{\rm B}$ denotes the thermally-averaged Gaunt factor, and its value depends on the energy of emitting electrons. We take $\bar{g}_{\rm B} = 1$ in our analysis \cite{dihingia-2020a, Sen-2022, Patra-2023}. The emissivity in  Eq.~(\ref{eq:emissivity}) is obtained with respect to an observer in the co-rotating frame. However, for a distant observer, this radiation gets red-shifted by the immense gravity of the central black hole. The emitted radiation also alters due to the rotation of the disc. Therefore, the expression of bolometric disc luminosity for a distant observer  is obtained as,
 \begin{equation}
 	\label{eq:bolometric-luminosity}	
 	L = \int_{0}^{\infty}L_{\nu_o}d\nu_{o}~{\rm erg~s^{-1}},
 \end{equation}	
 where
 \begin{equation}
 	\label{eq:monochromatic-luminosity}
 	L_{\nu_o} = 2\int_{r_0}^{r_{\rm edge}}\int_{0}^{2\pi}\mathcal{E}_{\nu_o}^{\rm ff}Hr~dr d\phi~{\rm erg~s^{-1}~Hz^{-1}}.
 \end{equation}	
Here $r_0 = r_{\rm H}$ is the disc inner edge, and $r_{\rm edge}$ is the disc outer edge. If the inclination angle of the accretion disc with respect to the distant observer is $\theta_0$, then the observed frequency $\nu_{o}$ is related with the emission frequency $\nu_e$ and the redshift factor ($1 + z$) as \cite{dihingia-2020a, Sen-2022},
\begin{equation}
	\label{eq:red-shift}
	\begin{split}
	\frac{\nu_e}{\nu_o}= 1 + z = u^{t}\left(1 + r\Omega\sin{\phi}\sin\theta_0\right),
	\end{split}
\end{equation}
where $\Omega = f_2/f_1$. In this work, we consider $\theta_0 = 45^{\circ}$ for illustration. The expression of $u^{t}$ is obtained using the time-like condition $u^{k}u_{k} = -1$ as, 
\begin{equation}
	\label{eq:u_upper_t}
	\begin{split}
		u^{t} = \gamma_{v}\sqrt{\frac{r^3}{(1 - \Omega \lambda)(\Omega a_{\rm k}(2r^2 + \eta_0) + r^3 - 2r^2 - \eta_0)}}.
	\end{split}
\end{equation}

The above equations are useful for finding the accretion solutions and their corresponding disc properties, such as temperature profile, radiation spectrum, luminosity, etc. In the subsequent sections, the results corresponding to the effect of deformation parameter on the accretion flow in KZ spacetime will be shown.

\section{Results}
\label{sec:result}
\subsection{Transonic accretion solutions}
\label{sec:accretion-solutions}

\begin{figure}
	\centering
	\includegraphics[width=\columnwidth]{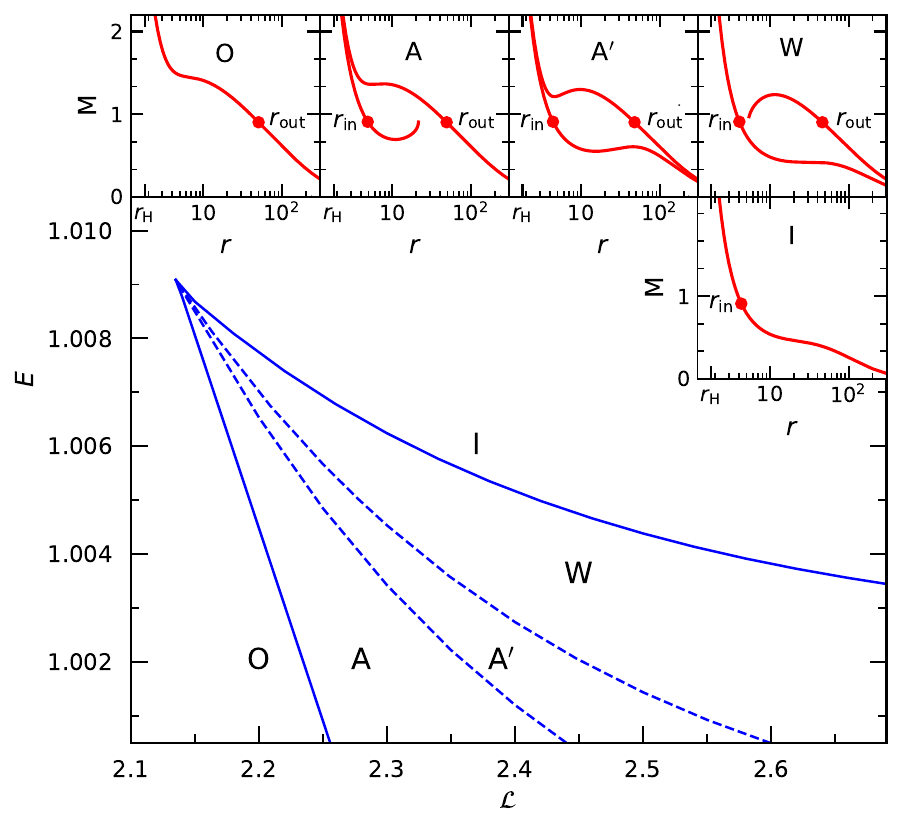}
	\caption{Classification of $\mathcal{L}-E$ parameter space into five regions (O, A, $\rm A^\prime$, W, and I) based on the accretion solution topologies. The inset panels represent their respective accretion solutions (Mach number ($M = v/C_s$) vs radial distance ($r$) curves). In this figure, we choose $a_{\rm k} = 0.65$, $\eta_0  = 0.1$, and $\alpha = 0.05$. See the text for details.}
	\label{fig:ps}
\end{figure}

In this section, we investigate the properties of accretion solutions and their associated parameter space in the KZ spacetime. For transonic accretion,  flow needs to pass a critical point, which is characterized by $dv/dr \rightarrow ``0/0"$ \cite{liang-1980, abramowicz-1981, lu-1985, Chakrabarti-1996}. Using the condition $\mathcal{N} = \mathcal{D} = 0$ and Eqs.~(\ref{eq:L}) and (\ref{eq:E}), the critical point ($r_c$) and corresponding flow variables ($\Theta_c, v_c, \lambda_c$) have been numerically determined for a given set of input parameters ($a_{\rm k}, \eta_0, \alpha, E, \mathcal{L}$). We use the l$'$H\^{o}pital's rule to calculate $dv/dr$ at $r_c$. Usually, it possesses two distinct values, and critical points are classified depending on them, e.g., saddle, nodal and spiral types. When both values of $(dv/dr)_{r_c}$ are real with opposite signs, the critical points are classified as saddle-types. For real and same sign of $(dv/dr)_{r_c}$ values gives the nodal-type critical points. In the case of spiral-type critical points, $(dv/dr)_{r_c}$ values are found to be imaginary. In our study, we are only interested in the accretion solutions that pass through the saddle-type critical points, as these are physically acceptable. The positive value of $(dv/dr)_{r_c}$ gives an accretion solution, and the negative value of $(dv/dr)_{r_c}$ corresponds to the wind solution. To find an accretion solution, we numerically solve the differential Eqs.~(\ref{eq:Temperature-gradient}), (\ref{eq:angular-momentum-gradient}), and (\ref{eq:velocity-gradient}) by utilizing the calculated parameters ($\Theta_c, v_c, \lambda_c$) at $r_c$ as the initial boundary condition. We first integrate the equations from $r_c$ to the disc inner edge $r_{0} = r_{\rm H}$, and then from $r_c$ to the disc outer edge $r_{\rm edge}$. Ultimately, by combining these two segments of the solutions, we obtain the resulting global transonic accretion solution for same set of input parameters ($a_{\rm k}, \eta_0, \alpha, E, \mathcal{L}$).

Following the above mentioned methodology, we investigate the structure of transonic accretion solutions and find the associated parameter space in the $\mathcal{L}-E$ plane. We obtain five distinct types of accretion solutions, and thereby, we subdivide the $\mathcal{L}-E$ parameter space into five designated regions as O, A, $\text{A}^{\prime}$, W, and I in Fig.~\ref{fig:ps}. Here, we choose $a_{\rm k} = 0.65$, $\eta_0 = 0.1$, and $\alpha = 0.05$. The accretion solutions ($\it i.e.,$ variation of Mach number $M = |v|/C_s$ as a function of $r$) for the selected parameters from the identified regions are displayed in the inset panels. The O-type solution is obtained for $(\mathcal{L}, E) = (2.2, 1.004)$, where the flow possesses only a outer critical point at $r_{\rm out} = 50.1375$, and the solution passing through it extends from $r_{\rm edge}$ to $r_{\rm H}$ (which is signature of a open solution). For A-type solution, we choose $(\mathcal{L}, E) = (2.25, 1.004)$, where the flows contain multiple critical points. In this case, the solution passing through $r_{\rm out} = 48.8889$ is a open solution. However, the solution passing through the inner critical point $r_{\rm in} = 4.8702$ terminates at $r_{t} = 21.4802$, which is referred to as a closed solution. In our study, we mainly focus on the open solutions since the closed solutions are physically unacceptable. However, in case of shock transition \cite{dihingia-2018, Dihingia-2019, Patra-2022, Sen-2022, Mitra-2024}, where flow connects the outer sonic point solution to the inner one, we deal with both open and closed solutions (see Section~\ref{sec:shock}). The results corresponding to the $\text{A}^{\prime}$-type solution is calculated for $(\mathcal{L}, E) = (2.3, 1.004)$. It is observed that the flow again exhibits multiple critical points ($r_{\rm in} = 4.3771, r_{\rm out} = 47.5241$). Interestingly, in this case, solutions passing through both critical points remain open, thereby leading to a degeneracy in accretion solutions. To remove the degeneracy, we calculate the entropy content $\dot{\mathcal{M}}$ (which is a constant of motion for adiabatic flow) very close to the event horizon ($r_{\rm H} = 1.7957$) using Eq.~(\ref{eq:entropy-accretion-rate}). At a chosen radial distance of $r = 1.8171$, the calculated values of $\dot{\mathcal{M}}$ are $2.9799 \times 10^8$ and $2.2219 \times 10^8$ for the inner and outer critical point solutions, respectively. Since the solution through $r_{\rm in}$ exhibits a higher entropy than the solution through $r_{\rm out}$, the previous one is considered a more naturally favorable than the latter one in accordance with the second law of thermodynamics, thereby resolving the degeneracy problem. For the W-type solution, we take $(\mathcal{L}, E) = (2.35, 1.004)$, where the flows still possess multiple critical points. In this scenario, the solution remains open through $r_{\rm in} = 4.0632$, while it becomes closed ($r_{t} = 5.4046$) through $r_{\rm out} = 46.0146$. Finally, the I-type accretion solution is calculated for $(\mathcal{L}, E) = (2.3, 1.0065)$, where the global solution only pass through $r_{\rm in} = 4.3054$. We point out that the accretion solutions depicted in inset panels are identical to the transonic solutions identified in Kerr spacetime \cite{Dihingia-2019}. Moreover, for non-viscous flow, such $\text{A}^{\prime}$-type solutions are not present, both in the Kerr and KZ non-Kerr spacetime. Hence, these particular solutions stem from the effect of viscosity only, not dealing with the spacetime deformation. It is essential to mention that there are no qualitative distinctions in the characteristics of accretion solutions for different $a_{\rm k}$ values other than reducing available parameter space when $a_{\rm k}$ increases. In this paper, all analyses have been conducted for moderately spinning black holes with $a_{\rm k} = 0.65$ to avail a fair range of ($\mathcal{L}, E$). Indeed, the above analysis intimates that the regions marked as O, A, $\text{A}^{\prime}$, W, and I provide characteristically different accretion solutions. However, the region A has a special interest in the realm of shock solutions, which will be studied in the next subsection.

\subsection{Effect of deformation on physical properties of accretion disc for shock solutions}
\label{sec:shock}

\begin{figure}
	\centering
	\includegraphics[width=\columnwidth]{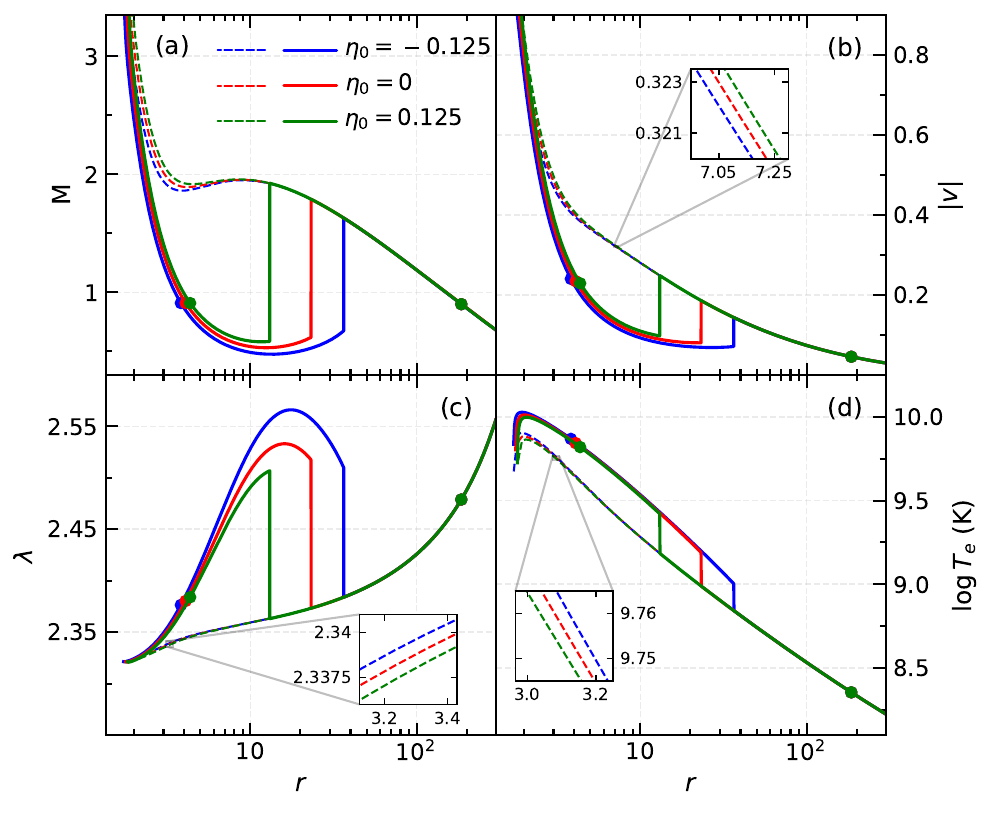}
	\caption{Typical shock induced accretion solutions (Mach number ($M$) vs radial distance ($r$) curves (solid) in panel (a)) along with their flow variables (radial velocity ($v$) in panel (b), specific angular momentum ($\lambda$) in panel (c), and electron temperature ($T_e$) in panel (d)) for deformation parameters $\eta_0 = -~0.125, 0$ and $0.125$. Here, the dashed curves represent the scenario where shock transitions have not occurred. The set of input parameters are chosen as $a_{\rm k} = 0.65$, $\alpha = 0.05$, $\mathcal{L} = 2.325$, and $E = 1.001$. See the text for details.}
	\label{fig:fp}
\end{figure}

Here, we examine accretion solutions capable of accommodating shock transitions in the deformed spacetime, which are preferred for their higher entropy content compared to the solutions without shocks \cite{Becker-2001}. In Fig.~\ref{fig:fp}a, we demonstrate typical shock solutions in accretion flow for different values of the deformation parameter ($\eta_0$). The blue, red, and green color solid curves represent the obtained results for $\eta_0 = -~0.125$, $0$, and $0.125$, respectively. We consider the flows start accretion at the disc outer edge with a given value of $(\mathcal{L}, E) = (2.325, 1.001)$. We also take $\alpha = 0.05$. The critical points associated with these shock solutions are summarized in Table~\ref{tab:table-1}. In a normal situation, it is observed that during accretion, the subsonic flows gradually achieve their supersonic states after passing through the outer critical point ($r_{\rm out}$) and continue to proceed towards the central black hole until they cross the event horizon, as shown by the dashed curves. However, due to the dominance of centrifugal repulsion over the gravitational pull, the incoming flow slows down, causing discontinuous jump into the flow variables in between $r_{\rm in}$ and $r_{\rm out}$. To find the shock location, we use the relativistic shock conditions which are expressed as \cite{Taub-1948},
\begin{itemize}
	\item[(a)] Mass flux conservation: $[\rho u^r]$,
	\item [(b)] Energy flux conservation: $[(e + p)u^{r}u^{t} + \pi^{rt}]$,
	\item [(c)] Radial-momentum flux conservation:\\ $[(e + p)u^{r}u^{r} + \pi^{rr}]$.
\end{itemize}
Here, the square brackets denote the difference between the pre-shock and post-shock quantities. The explicit form of the stress tensor component $\pi^{rr}$ is calculated using Eq.~(\ref{eq:shear-tensor}), and has been given in Appendix \ref{appendix-A}. For $\eta_0 = -~0.125$, $0$, and $0.125$, we obtain the shock locations at $r_{\rm sh} = 36.4466$, $23.1665$, and $13.0675$, respectively, shown by the solid vertical lines. After the shock transition from the supersonic to the subsonic branch, flow again becomes supersonic when it passes through the inner critical point ($r_{\rm in}$), and continues to accrete up to the horizon. Therefore, in the shock scenario, accretion solution can passes through both $r_{\rm in}$ and $r_{\rm out}$. Here, it is observed that the shock locations move closer to the horizon as the deformation parameter is increased. An explanation for this will be provided later within this section.

\begin{table}
	\centering
	\caption{Deformation parameter ($\eta_0$), event horizon location ($r_{\rm H}$), critical point locations ($r_{\rm in}, r_{\rm out}$), shock location ($r_{\rm sh}$) for shock-induced global accretion solutions presented in Fig. \ref{fig:fp}.}
	\label{tab:table-1}
	\begin{ruledtabular}
		\begin{tabular}{lcccc}
			$\eta_0$ & $r_{\rm H}$ & $r_{\rm in}$ & $r_{\rm out}$ & $r_{\rm sh}$ \\
			\hline
			-~0.125 & 1.7102 & 3.8166 & 185.8394 & 36.4466\\
			0 & 1.7599 & 4.0765 & 185.8436 & 23.1665\\
			0.125 & 1.8042 & 4.3375 & 185.8478 & 13.0675\\
		\end{tabular}
	\end{ruledtabular}
\end{table}

 As there are discontinuous jumps in the flow variables during shock transitions, it is therefore worthy to study them across the shock fronts. We examine different physical properties of the accretion disc associated with the shock-induced accretion solutions in the deformed spacetime. In Fig.~\ref{fig:fp}b-d, we present the profiles of radial 3-velocity ($v$), specific angular momentum ($\lambda$), and electron temperature ($T_e$) corresponding to the accretion solutions in Fig.~\ref{fig:fp}a, where the sudden jumps in the analyzed quantities are clearly visible for the shock solutions. Furthermore, it is observed that the difference of these quantities across the shock front increases when the deformation parameter increases, which is expected as the shock fronts settle down at smaller radii. As the deformation parameter increases, causing the event horizon radius to expand, the critical points shift towards larger radii (see Table~\ref{tab:table-1}). It increases the radial velocity of flow, and consequently decreases the specific angular momentum (see panels (b) and (c)). Therefore, the increase of $\eta_0$ weakens the centrifugal repulsion against gravity, ultimately pushing the shock locations at smaller radii. Since the radial velocity drops down at the shock radius, the kinetic energy of pre-shock flow converted into the thermal energy, and consequently increase the electron-temperature of post-shock flow (equivalently PSC). It is noteworthy that the soft photons from the pre-shock disc may inverse Comptonize by the hot electrons in PSC, and produce the high energy radiation in BH-XRBs \cite{majumder-2022, nandi-2024, Chatterjee-2024}.

\begin{figure}
	\centering
	\includegraphics[width=\columnwidth]{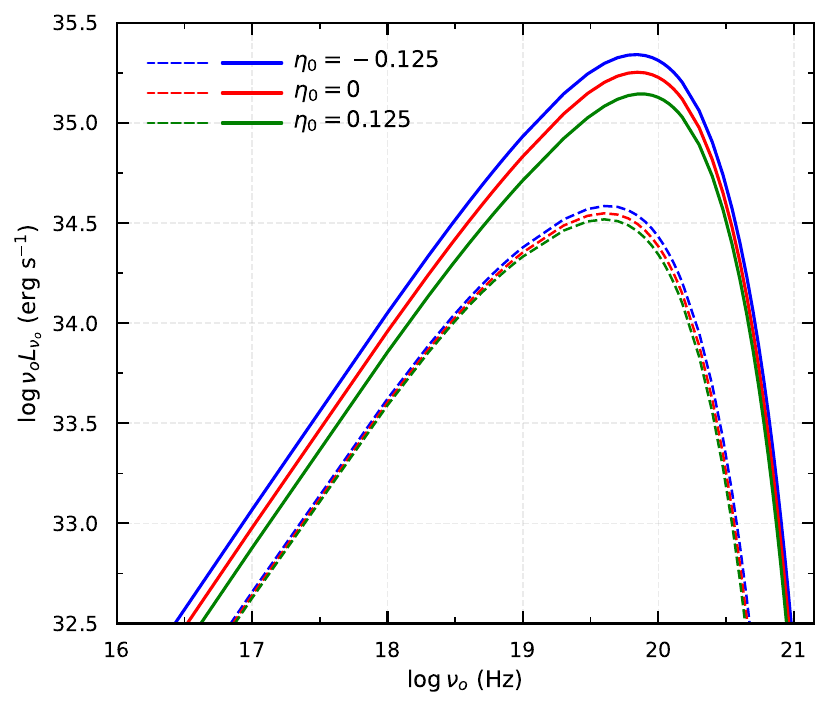}
	\caption{Spectral luminosity distributions of radiation emitted from the accretion disc corresponding to the same set of accretion solutions in Fig.~\ref{fig:fp}. Here, the input parameters are chosen as $r_{\rm edge} = 300$, $M_{\rm BH} = 10M_{\odot}$, and $\dot{M} = 0.1\dot{M}_{\rm Edd}$. See the text for details.}
	\label{fig:SED}
\end{figure}

Now, we examine the disc spectrum of emitted radiation from the accretion disc for the shock-free and shock-induced solutions. We calculate the spectral energy distributions (SEDs) associated with the accretion solutions of Fig.~\ref{fig:fp}a using Eq.~(\ref{eq:monochromatic-luminosity}). In this work, we consider $r_{\rm edge} = 300$, $M_{\rm BH} = 10M_{\odot}$, and $\dot{M} = 0.1\dot{M}_{\rm Edd}$, where $\dot{M}_{\rm Edd} = 1.39 \times 10^{18}M_{\rm BH}/M_{\odot}~\rm gm~s^{-1}$ is the  Eddington mass accretion rate. The obtained results are depicted in Fig.~\ref{fig:SED}, illustrating the variation of quantity $\nu_{o}L_{\nu_o}$ as a function of the observed frequency $\nu_{o}$. Note that the bremsstrahlung radiation maximizes its power at $\nu_o \approx 10^{20}{\rm Hz}$, and cut-off at $\nu_o \approx 10^{21}{\rm Hz}$ ($= k_{\rm B}T_{e0}/h$, which is corresponds to the disc inner edge electron temperature $T_{e0} \approx 10^{10}\text{K}$). We see that the shock-induced solutions produce much higher SEDs than the shock-free solutions. This happens because the temperature of a shock solution is larger than that of a shock-free solution. Furthermore, we observe that increasing $\eta_0$ reduces the luminosity distributions due to the decrease in temperature of the accretion disc (see Fig.~\ref{fig:fp}e).

\begin{figure}
	\centering
	\includegraphics[width=\columnwidth]{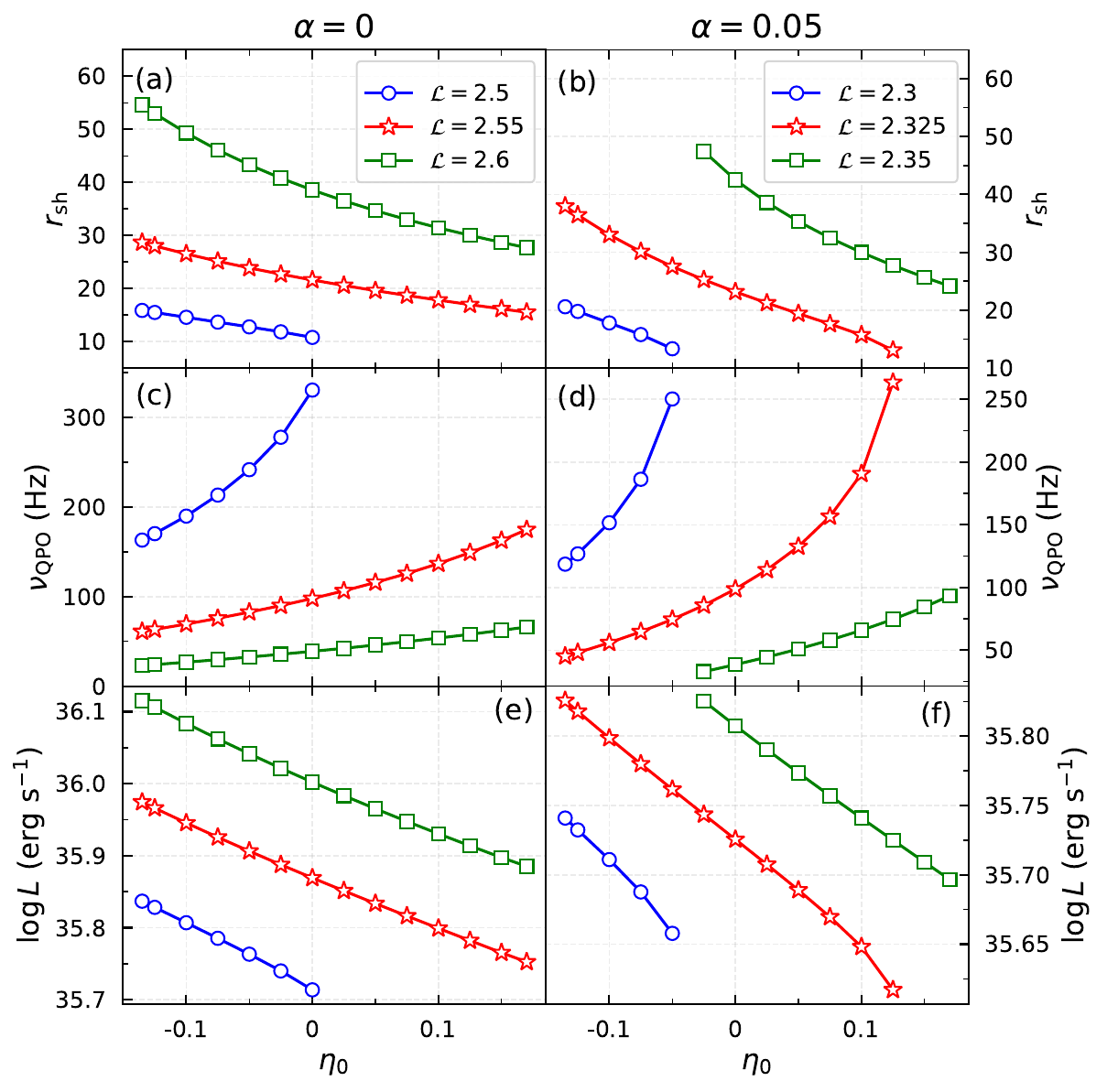}
	\caption{Variation of shock location ($r_{\rm sh}$), QPO frequency ($\nu_{\rm QPO}$), and bolometric disc luminosity ($L$) with the deformation parameter ($\eta_0$) for $\alpha = 0$ (left panels) and for $\alpha = 0.05$ (right panels). In this figure, the set of input parameters are taken as $a_{\rm k} = 0.65$, $E = 1.001$, $r_{\rm edge} = 300$, $M_{\rm BH} = 10M_{\odot}$, and $\dot{M} = 0.1\dot{M}_{\rm Edd}$. The bulk angular momentum $\mathcal{L}$ is marked in each panel. See the text for details.}
	\label{fig:sp}
\end{figure}

Next, we investigate the effect of deformation parameter on various shock properties, and the obtained results are presented in Fig.~\ref{fig:sp} for flow energy $E = 1.001$. In panel (a), we show the variation of shock location ($r_{\rm sh}$) as a function of $\eta_0$ for a non-viscous flow (${\it i.e.,}~\alpha = 0$), where the open circles (blue), open asterisks (red), and open squares (green) joined with the solid lines denote the results for $\mathcal{L} = 2.5, 2.55$, and $2.6$, respectively. Similarly, in panel (b), we present the variation of $r_{\rm sh}$ for a viscous flow with $\alpha = 0.05$, where the results corresponding to $\mathcal{L} = 2.3, 2.325$, and $2.35$ are plotted using the open circles (blue), open asterisks (red), and open squares (green) joined with the solid lines. In both panels, we notice that $r_{\rm sh}$ decreases with the increase of $\eta_0$, irrespective of $\mathcal{L}$ values. These findings are in agreement with the results of Fig.~\ref{fig:fp}. Notably, the deformation parameter cannot be increased indefinitely because the shock conditions cease to be satisfied beyond a critical value ($\eta_0 > \eta_{0}^{\rm cri}$), leading to the disappearance of shock transitions. From numerical simulation \cite{molteni-1995, Chakrabarti-2015, Kim-2018}, we observe that the shock can oscillate along the radial direction in a quasi-periodic manner when the infall time from the shock location is comparable to the radiative cooling time of post-shock flow, which leads to the oscillation of hard X-rays during the outburst phases of black hole binaries (BHBs) and black hole candidates (BHCs). The frequency of quasi-periodic oscillation (QPO) of the shock front is expressed as $\nu_{\rm QPO} \approx 1/t_{\rm infall}$ \cite{molteni-1995}, where $t_{\rm infall} = \int_{r_{\rm sh}}^{r_{\rm H}}v^{-1}dr$ is the infall time of post-shock flow. We calculate $\nu_{\rm QPO}$ corresponding to the result of panels (a) and (b) by assuming that these shocks are oscillatory. The obtained results are depicted in the respective panels (c) and (d), where $\nu_{\rm QPO}$ is plotted as a function of $\eta_0$. We observe that $\nu_{\rm QPO}$ increases with the increase of $\eta_0$ for both the viscous and non-viscous flows. As shocks settle down at smaller radii for higher $\eta_0$ values, $t_{\rm infall}$ decreases, which gives high $\nu_{\rm QPO}$ values. It is noteworthy that the theoretically calculated QPO frequencies for a $10M_{\odot}$ black hole are consistent with the observed QPOs for BHBs and BHCs \cite{Remillard-2006a}. Finally, we calculate the bolometric disc luminosity ($L$) using Eq.~(\ref{eq:bolometric-luminosity}) for the global shock solutions of panels (a) and (b). We present the obtained results in panels (e) and (f), respectively, where we notice that $L$ decreases in both cases with an increase in $\eta_0$. This occurs because enhanced deformation reduces the luminosity distribution in the emitted radiation (see Fig.~\ref{fig:SED}), and consequently decreasing disc luminosity. 

\section{Phenomenology on high frequency quasi-periodic oscillations}
\label{sec:HFQPO}

\begin{figure}
	\centering
	\includegraphics[width=\columnwidth]{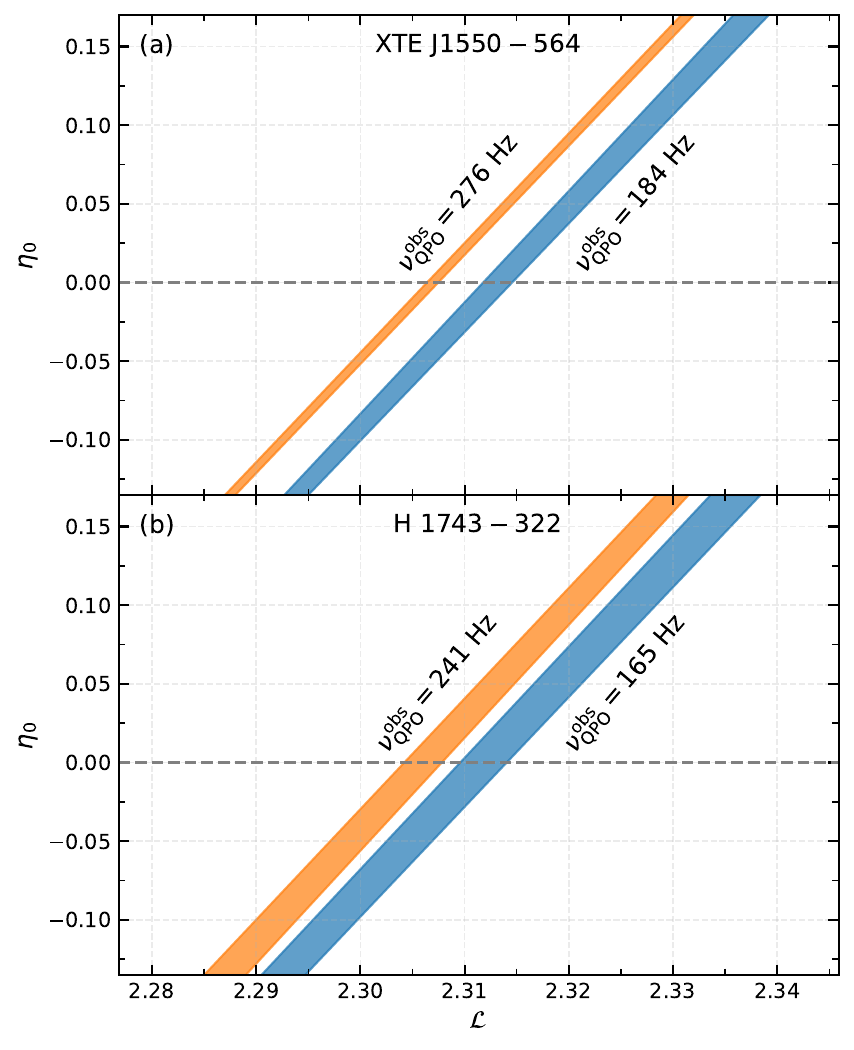}
	\caption{Model estimated parameter space of deformation parameter ($\eta_0$) and bulk angular momentum ($\mathcal{L}$) for observed QPO frequencies in black hole microquasars ${\rm XTE}~{\rm J}1550-564$ and ${\rm H}~1743-322$. The horizontal dashed lines (gray) represent the results corresponding to the Kerr black holes. In this figure, we take $M_{\rm BH} = (9.1 \pm 0.61) M_{\odot}$ for ${\rm XTE}~{\rm J}1550-564$, and $M_{\rm BH} = 11.21^{+1.65}_{-1.96} M_{\odot}$ for ${\rm H}~1743-322$. Here, other input parameters are chosen as $a_{\rm k} = 0.65$, $\alpha = 0.05$, and $E = 1.001$. See the text for details.}
	\label{fig:L-eta}
\end{figure}

In this section, we find the parameter space of bulk angular momentum ($\mathcal{L}$) and deformation parameter ($\eta_0$) corresponding to the observed QPO frequencies ($\nu_{\rm QPO}^{\rm obs}$) in X-ray spectra of some BHBs and BHCs. For that, we choose two well-studied sources, namely ${\rm XTE}~{\rm J}1550-564$ and ${\rm H}~1743-322$. These black hole microquasars are commonly known for their pairs of high-frequency QPOs (HFQPOs) with moderate black hole spin values. For ${\rm XTE}~{\rm J}1550-564$, the observed pairs of QPO frequencies are $184, 276~{\rm Hz}$ \cite{Homan-2000, Miller-2001, Remillard-2002}, and for ${\rm H}~1743-322$, those values are $165, 241~{\rm Hz}$ \cite{Homan-2005, Remillard-2006b}. The reported mass of black hole primary of BHB ${\rm XTE}~{\rm J}1550-564$ is $M_{\rm BH} = (9.1 \pm 0.61)M_{\odot}$ \cite{Orosz-2011}, and the black hole mass of BHC ${\rm H}~1743-322$ is provided as $M_{\rm BH} = 11.21^{+1.65}_{-1.96} M_{\odot}$ \cite{Molla-2016}. Also, the estimated spin parameters of black holes via continuum-fitting method (CFM) are $-0.11 < a_{\rm k} < 0.71$ \cite{Steiner-2011} for ${\rm XTE}~{\rm J}1550-564$, and $-0.3 < a_{\rm k} < 0.7$ \cite{Steiner-2012} for ${\rm H}~1743-322$. In this work, we consider $a_{\rm k} = 0.65$ for both sources for the purpose of illustration. We further take the other input parameters as $\alpha = 0.05$ and $E = 1.001$. Using the above mentioned parameters for ${\rm XTE}~{\rm J}1550-564$ source, we calculate the parametric regions of $\mathcal{L}$ and $\eta_0$ in Fig.~\ref{fig:L-eta}(a), where the shaded region (in blue) is obtained for $\nu_{\rm QPO}^{\rm obs} = 184~{\rm Hz}$, and the shaded region (in orange) is depicted for $\nu_{\rm QPO}^{\rm obs} = 276~{\rm Hz}$. Similarly, in Fig.~\ref{fig:L-eta}(b), we present the obtained results in $\mathcal{L}-\eta_0$ plane for ${\rm H}~1743-322$, where the blue and orange shaded regions correspond to $\nu_{\rm QPO}^{\rm obs} = 165, 241~{\rm Hz}$, respectively. In both panels, we observe that the effective region of the parameter space is shifted towards the lower $\mathcal{L}$ domain when $\nu_{\rm QPO}^{\rm obs}$ is increased. Moreover, we notice that for a given $\mathcal{L}$, $\eta_0$ takes higher values corresponding to a larger $\nu_{\rm QPO}^{\rm obs}$ than the smaller $\nu_{\rm QPO}^{\rm obs}$. This happens because, in order to produce high QPO frequencies, the shock front has to move towards the inner edge of the disc, which is achieved by reducing angular momentum and increasing deformation (see Figs.~\ref{fig:sp}a, b). We further notice that the area under parameter space reduces with the increase of $\nu_{\rm QPO}^{\rm obs}$ for both scenarios.

\section{Conclusions}
\label{sec:conclusion}
In this work, we study the viscous accretion flow around the KZ deformed Kerr black hole in the light of general relativistic hydrodynamics. We construct a set of equations that govern the dynamics of accretion flow in the disc. These equations have been solved numerically, and we are able to find the accretion solutions that satisfy the boundary conditions for the transonic flow. We explore all possible types of accretion solutions and find that these solutions are not unique; instead, a parameter space exists. Since the deformation in KZ spacetime alters the near horizon geometry, a key aspect of our investigation is to analyze the impact of deformation parameter ($\eta_0$) on the physical properties of accretion disc. Accordingly, we explore the radiation spectrum of accretion disc as a function of $\eta_0$. We find that as $\eta_0$ increases, the luminosity distribution decreases in both the shock-free and as well as the shock-induced accretion solutions. Further, we make an effort to examine the effect of deformation on the shock properties, mainly the shock location ($r_{\rm sh}$), QPO frequency ($\nu_{\rm QPO}$), and bolometric disc luminosity ($L$). We also find that the shock fronts settle at smaller radii for higher $\eta_0$ values, which increases $\nu_{\rm QPO}$ and decreases $L$. Hence, the present study indicates that $\eta_0$ significantly influences various physical properties of the accretion disc.

Further, we phenomenologically identify the $\mathcal{L}-\eta_0$ parameter spaces associated with the pairs of HFQPOs observed in ${\rm XTE}~{\rm J}1550-564$ and ${\rm H}~1743-322$ sources. Here, we find that when the observed QPO frequency increases, the obtained parameter space moves towards the lower $\mathcal{L}$ side and shrinks as well. Most importantly, using this phenomenological approach, the identified non-zero $\eta_0$ values (which offer parametric deviations to the Kerr metric) indicate that the KZ non-Kerr black holes can also describe the observed spectral properties of the accretion disc, like the Kerr black holes in Einstein's gravity theory.

Now, we provide a comparative discussion for the accretion properties corresponding to the present type of deformation (i.e.~the KZ case) and the JP type. In our earlier study of accretion flow in the JP spacetime \cite{Patra-2022, Patra-2023}, we self-consistently found all possible classes of the transonic accretion solutions, including the shock solutions, using the GR hydrodynamics framework and the relativistic EoS. We mainly observed there the effect of deformation on the properties of accretion flows. We also noticed that as the JP deformation increases, critical points are formed at the smaller radii due to a decrease in the event horizon radius. Consequently, the shock front recedes away from the horizon, resulting in a decrease in the change of flow variables such as density, temperature, etc., across the shock fronts. Moreover, in the spectral analysis of the accretion disc, we demonstrated that the luminosity distribution of emitted radiation increases with increase of the JP deformation. Based on these prior analyses, we see that the accretion solutions in the KZ spacetime are characteristically similar to those in the JP spacetime. Hence, by looking at the behavior of accretion solutions, it is difficult to isolate them. However, by studying various disc properties (e.g., shock radii, luminosity distributions, etc.) corresponding to a particular class of accretion solutions in these two spacetime, we observe that the KZ deformation shows opposite behavior in controlling the disc properties from the JP deformation. Therefore, we wish to emphasize that the analysis of transonic accretion flows around various {kind} of non-Kerr black holes serves as an alternative theoretical approach to characterize the parametric deviations from the Kerr metric.

Finally, we mention the limitations of this work. Our study does not include the radiative cooling \cite{Sarkar-2018, Sarkar-2022}, thermal conduction \cite{Mitra-2023, Singh-2024-02256}, magnetic fields \cite{Mitra-2022, Mitra-2024}, etc., despite their inevitable presence within the accretion disc. We plan to explore these aspects in the future work and report our findings separately.

\section*{Data availability statement}
The data underlying this article will be available with reasonable request.

\section*{Acknowledgments}
Authors thank the anonymous reviewer for constructive comments and useful suggestions that help to improve the quality of the paper. SP thanks Indu Kalpa Dihingia, Sumit Dey and Seshadri Majumder for useful discussions. The work of SP is supported by the University Grants Commission (UGC), Government of India, under the scheme Senior Research Fellowship (SRF). BRM is supported by a START-UP RESEARCH GRANT (No. SG/PHY/P/BRM/01) from the Indian Institute of Technology Guwahati (IITG), India. The work of SD is supported by the Science and Engineering Research Board (SERB), India, through grant MTR/2020/000331. 
 
 %%%%%%%%%%%%%%%%%%%%%%%%%%%%%%%%%%%%%%%%%%%%%%%%%%%%%%%
\appendix
\begin{widetext}
	\section{Shear tensor components $\sigma^{r}_{\phi}$ and $\sigma^{r}_{t}$}
	\label{appendix-A}
	The $r-\phi$ and $r-t$ components of the viscous shear tensor ($\sigma^{i}_{k}$) are calculated as,
	\begin{equation*}
		\begin{split}
			& \sigma^{r}_{\phi} = \frac{1}{2}\left(A_{1} + A_{2}\frac{dv}{dr} + A_{3}\frac{d\lambda}{dr}\right),\\ 
			& \sigma^{r}_{t} = \frac{1}{2}\left(B_{1} + B_{2}\frac{dv}{dr} + B_{3}\frac{d\lambda}{dr}\right),
		\end{split}
	\end{equation*}
	where
	\begin{equation*}
		\begin{split}
			& A_1 = - \frac{u_t\gamma_{v}^{2}(A_{11} + A_{12})}{r^3},~A_{11} = v^2(\Delta r - \eta_0)(\lambda P_2 - P_1),~A_{12} = \frac{2\lambda L_N + 2\mathcal{G}(\Delta r - \eta_0)P_1 - \lambda (\Delta r - \eta_0)^2 P_3}{2\mathcal{G}},\\
			& A_2 = - \frac{4\lambda u_t v\gamma_{v}^{4}(\Delta r - \eta_0)}{3r^{3}},~A_{3} = - \frac{u_t\gamma_{v}^2(\Delta r - \eta_0)\left(r^2(r^3 + a_{\rm k}^2(r + 2)) - a_{\rm k}\lambda(2r^2 + \eta_0) + a_{\rm k}^2\eta_0\right)}{\mathcal{G}r^5},\\
			& B_1 = \frac{u_t\gamma_{v}^{2}(B_{11} +B_{12})}{r^3},~B_{11} = v^2(\Delta r - \eta_0)(Q_1 + Q_2),~B_{12} = \frac{2L_N - 2\mathcal{G}(\Delta r - \eta_0)Q_1 - (\Delta r - \eta_0)^2Q_3}{2\mathcal{G}},\\
			& B_2 = \frac{4u_t v\gamma_{v}^{4}(\Delta r - \eta_0)}{3r^3},~B_{3} = \frac{u_t\gamma_{v}^2(\Delta r - \eta_0)(a_{\rm k}(2r^2 + \eta_0) + \lambda(r^3 - 2r^2 - \eta_0))}{\mathcal{G}r^5},\\
			& u_t = - \frac{\gamma_{v}}{\sqrt{\lambda f_2 - f_1}},~\mathcal{G} = \frac{r^2(r^3 + a_{\rm k}^2(r + 2)) - 2a_{\rm k}\lambda(2r^2 + \eta_0) - \lambda^2(r^3 - 2r^2 - \eta_0) + a_{\rm k}^2\eta_0}{r^2},\\ 
			& L_N = \frac{\mathcal{G}(\Delta + 2r(r - 1)) - (\Delta r - \eta_0)\mathcal{G}^\prime}{2},~\mathcal{G}^\prime = \left(\frac{d\mathcal{G}}{dr}\right)_{\Theta, v, \lambda},~P_1 = \frac{f_{1}a_{\rm k}(2r^{2} + 3\eta_0) + f_{2}(2r^{2}(r^3 - a_{\rm k}^2) - 3a_{\rm k}^{2}\eta_0)}{r^4},\\
			& P_2 = \frac{\eta_0 - 2r(a_{\rm k}^{2} + r(2r - 3))}{3r(\Delta r - \eta_0)},~P_3 = \frac{- f_{1}^{2}(2r^2 + 3\eta_0) + 2a_{\rm k}f_{1}f_{2}(2r^2 + 3\eta_0) + f_{2}^{2}(2r^{2}(r^3 - a_{\rm k}^2) - 3a_{\rm k}^{2}\eta_0)}{r^4},\\ 
			& Q_1 = \frac{(2r^2 + 3\eta_0)(a_{\rm k}f_2 - f_1)}{r^4},~Q_2 = P_2,~Q_3 = P_3.
		\end{split}
	\end{equation*}
	The $r-r$ component of the viscous stress tensor ($\pi^{ik}$) is given by,
	\begin{equation*}
		\begin{split}
		 & \pi^{rr} = -\frac{\eta (\Delta r - \eta_0)}{r^3}\left(C_{1} + C_{2}\frac{dv}{dr}\right),\\
		 & C_1 = \frac{v\gamma_{v}}{r}\sqrt{\frac{\Delta r - \eta_0}{r}}\left(\frac{\gamma_{v}^{2}(2r(a_{\rm k}^2 - r) - 3\eta_0)}{3r(\Delta r - \eta_0)} - \frac{4\gamma_{v}^{2}}{3r} - u_{t}^{2}P_3\right),~C_2 = \frac{4\gamma_{v}^5}{3r}\sqrt{\frac{\Delta r - \eta_0}{r}}.
		\end{split}
	\end{equation*}
	Here, all the quantities have their usual meaning.
	
	\section{Radial-momentum equation, and conservation equations of $\mathcal{L}$ and $E$}
	\label{appendix-B}
	The conservation of bulk angular momentum ($\mathcal{L}$), conservation of Bernoulli constant ($E$), and r-component of momentum equation are respectively obtained as,  
	\begin{equation*}
		\begin{split}
			& \frac{d\mathcal{L}}{dr} = \mathcal{L}_{0} + \mathcal{L}_{1}\frac{dv}{dr} + \mathcal{L}_{2}\frac{d\Theta}{dr} + \mathcal{L}_{3}\frac{d\lambda}{dr} = 0,\\
			& \frac{dE}{dr} = E_{0} + E_{1}\frac{dv}{dr} + E_{2}\frac{d\Theta}{dr} + E_{3}\frac{d\lambda}{dr} = 0,\\
			& R_{0} + R_{1}\frac{dv}{dr} + R_{2}\frac{d\Theta}{dr} + R_{3}\frac{d\lambda}{dr} = 0,
		\end{split}
	\end{equation*}
	where
	\begin{equation*}
	\begin{split}
	& \mathcal{L}_0 = \lambda h_0 \left(\frac{du_t}{dr}\right)_{\Theta, v, \lambda} + \left( \frac{d(\Lambda A_1)}{dr}\right)_{\Theta, v, \lambda},\\
	& \mathcal{L}_2 = \frac{2\lambda u_t \Gamma}{(\Gamma - 1)(1 + m_p/m_e)} + \left( \frac{d(\Lambda A_1)}{d\Theta}\right)_{r, v, \lambda},\\
	& E_0 = h_0 \left(\frac{du_t}{dr}\right)_{\Theta, v, \lambda} - \left( \frac{d(\Lambda B_1)}{dr}\right)_{\Theta, v, \lambda},\\
	& E_2 = \frac{2 u_t \Gamma}{(\Gamma - 1)(1 + m_p/m_e)} - \left( \frac{d(\Lambda B_1)}{d\Theta}\right)_{r, v, \lambda},
	\end{split}
	\hspace{0.3cm}
	\begin{split}
	& \mathcal{L}_1 = \lambda h_0 \left(\frac{du_t}{dv}\right)_{r, \Theta, \lambda} + \left( \frac{d(\Lambda A_1)}{dv}\right)_{r, \Theta, \lambda},\\
	& \mathcal{L}_3 =  h_0u_t + h_0\lambda \left(\frac{du_t}{d\lambda}\right)_{r, \Theta, v} + \left( \frac{d(\Lambda A_1)}{d\lambda}\right)_{r, \Theta, v},\\
	& E_1 = h_0 \left(\frac{du_t}{dv}\right)_{r, \Theta, \lambda} - \left( \frac{d(\Lambda B_1)}{dv}\right)_{r, \Theta, \lambda},\\
	& E_3 = h_0 \left(\frac{du_t}{d\lambda}\right)_{r, \Theta, v} - \left( \frac{d(\Lambda B_1)}{d\lambda}\right)_{r, \Theta, v},
	\end{split} 
	\end{equation*}	
	\begin{equation*}
	\begin{split}
	& \Lambda = \frac{\alpha C_s Hr}{v\gamma_v \sqrt{(\Delta r - \eta_0)/r}},~R_0 = \frac{\Theta}{(f + 2\Theta)}\left(- \frac{3}{r} + \frac{F_1^\prime}{F_1} + \frac{\lambda \Omega^\prime}{1 - \Omega \lambda} - \frac{2r^2(r - 1) + \eta_0}{r(\Delta r - \eta_0)}\right) + \left(\frac{d\Phi^{\rm eff}}{dr}\right)_{\lambda},\\
	& R_1 = \gamma_{v}^2\left(v - \frac{2\Theta}{(f + 2\Theta)v}\right),~R_2 = \frac{1}{f + 2\Theta},~R_3 = \frac{\Theta}{(f + 2\Theta)(1 - \Omega \lambda)} \left( \frac{d(\Omega \lambda)}{d\lambda}\right)_{r, \Theta, v}.
	\end{split} 
	\end{equation*}	
	Here, all the quantities have their usual meaning.
\end{widetext}	
%%%%%%%%%%%%%%%%%%%%%%%%%%%%%%%%%%%%%%%%%%%%%%%%%%%%%%%
\bibliographystyle{apsrev}
\bibliography{references} 

\begin{thebibliography}{94}
\expandafter\ifx\csname natexlab\endcsname\relax\def\natexlab#1{#1}\fi
\expandafter\ifx\csname bibnamefont\endcsname\relax
  \def\bibnamefont#1{#1}\fi
\expandafter\ifx\csname bibfnamefont\endcsname\relax
  \def\bibfnamefont#1{#1}\fi
\expandafter\ifx\csname citenamefont\endcsname\relax
  \def\citenamefont#1{#1}\fi
\expandafter\ifx\csname url\endcsname\relax
  \def\url#1{\texttt{#1}}\fi
\expandafter\ifx\csname urlprefix\endcsname\relax\def\urlprefix{URL }\fi
\providecommand{\bibinfo}[2]{#2}
\providecommand{\eprint}[2][]{\url{#2}}

\bibitem[{\citenamefont{Sullivan et~al.}(2020)\citenamefont{Sullivan, Yunes,
  and Sotiriou}}]{Sullivan-2019}
\bibinfo{author}{\bibfnamefont{A.}~\bibnamefont{Sullivan}},
  \bibinfo{author}{\bibfnamefont{N.}~\bibnamefont{Yunes}}, \bibnamefont{and}
  \bibinfo{author}{\bibfnamefont{T.~P.} \bibnamefont{Sotiriou}},
  \bibinfo{journal}{Phys. Rev. D} \textbf{\bibinfo{volume}{101}},
  \bibinfo{pages}{044024} (\bibinfo{year}{2020}), \eprint{1903.02624}.

\bibitem[{\citenamefont{Deser and Tekin}(2003)}]{Deser-2002}
\bibinfo{author}{\bibfnamefont{S.}~\bibnamefont{Deser}} \bibnamefont{and}
  \bibinfo{author}{\bibfnamefont{B.}~\bibnamefont{Tekin}},
  \bibinfo{journal}{Phys. Rev. D} \textbf{\bibinfo{volume}{67}},
  \bibinfo{pages}{084009} (\bibinfo{year}{2003}), \eprint{hep-th/0212292}.

\bibitem[{\citenamefont{Maartens and Koyama}(2010)}]{Maartens-2010}
\bibinfo{author}{\bibfnamefont{R.}~\bibnamefont{Maartens}} \bibnamefont{and}
  \bibinfo{author}{\bibfnamefont{K.}~\bibnamefont{Koyama}},
  \bibinfo{journal}{Living Rev. Rel.} \textbf{\bibinfo{volume}{13}},
  \bibinfo{pages}{5} (\bibinfo{year}{2010}), \eprint{1004.3962}.

\bibitem[{\citenamefont{Nampalliwar et~al.}(2020)\citenamefont{Nampalliwar,
  Xin, Srivastava, Abdikamalov, Ayzenberg, Bambi, Dauser, Garcia, and
  Tripathi}}]{Nampalliwar-2020}
\bibinfo{author}{\bibfnamefont{S.}~\bibnamefont{Nampalliwar}},
  \bibinfo{author}{\bibfnamefont{S.}~\bibnamefont{Xin}},
  \bibinfo{author}{\bibfnamefont{S.}~\bibnamefont{Srivastava}},
  \bibinfo{author}{\bibfnamefont{A.~B.} \bibnamefont{Abdikamalov}},
  \bibinfo{author}{\bibfnamefont{D.}~\bibnamefont{Ayzenberg}},
  \bibinfo{author}{\bibfnamefont{C.}~\bibnamefont{Bambi}},
  \bibinfo{author}{\bibfnamefont{T.}~\bibnamefont{Dauser}},
  \bibinfo{author}{\bibfnamefont{J.~A.} \bibnamefont{Garcia}},
  \bibnamefont{and} \bibinfo{author}{\bibfnamefont{A.}~\bibnamefont{Tripathi}},
  \bibinfo{journal}{Phys. Rev. D} \textbf{\bibinfo{volume}{102}},
  \bibinfo{pages}{124071} (\bibinfo{year}{2020}), \eprint{1903.12119}.

\bibitem[{\citenamefont{Carter}(1971)}]{Carter-1971}
\bibinfo{author}{\bibfnamefont{B.}~\bibnamefont{Carter}},
  \bibinfo{journal}{Phys. Rev. Lett.} \textbf{\bibinfo{volume}{26}},
  \bibinfo{pages}{331} (\bibinfo{year}{1971}).

\bibitem[{\citenamefont{Robinson}(1975)}]{Robinson-1975}
\bibinfo{author}{\bibfnamefont{D.~C.} \bibnamefont{Robinson}},
  \bibinfo{journal}{Phys. Rev. Lett.} \textbf{\bibinfo{volume}{34}},
  \bibinfo{pages}{905} (\bibinfo{year}{1975}).

\bibitem[{\citenamefont{Sen}(1992)}]{Sen-1992}
\bibinfo{author}{\bibfnamefont{A.}~\bibnamefont{Sen}}, \bibinfo{journal}{Phys.
  Rev. Lett.} \textbf{\bibinfo{volume}{69}}, \bibinfo{pages}{1006}
  (\bibinfo{year}{1992}), \eprint{hep-th/9204046}.

\bibitem[{\citenamefont{Kanti et~al.}(1996)\citenamefont{Kanti, Mavromatos,
  Rizos, Tamvakis, and Winstanley}}]{Kanti-1995}
\bibinfo{author}{\bibfnamefont{P.}~\bibnamefont{Kanti}},
  \bibinfo{author}{\bibfnamefont{N.~E.} \bibnamefont{Mavromatos}},
  \bibinfo{author}{\bibfnamefont{J.}~\bibnamefont{Rizos}},
  \bibinfo{author}{\bibfnamefont{K.}~\bibnamefont{Tamvakis}}, \bibnamefont{and}
  \bibinfo{author}{\bibfnamefont{E.}~\bibnamefont{Winstanley}},
  \bibinfo{journal}{Phys. Rev. D} \textbf{\bibinfo{volume}{54}},
  \bibinfo{pages}{5049} (\bibinfo{year}{1996}), \eprint{hep-th/9511071}.

\bibitem[{\citenamefont{Yunes and Pretorius}(2009)}]{Yunes-2009-084043}
\bibinfo{author}{\bibfnamefont{N.}~\bibnamefont{Yunes}} \bibnamefont{and}
  \bibinfo{author}{\bibfnamefont{F.}~\bibnamefont{Pretorius}},
  \bibinfo{journal}{Phys. Rev. D} \textbf{\bibinfo{volume}{79}},
  \bibinfo{pages}{084043} (\bibinfo{year}{2009}), \eprint{0902.4669}.

\bibitem[{\citenamefont{Johannsen and Psaltis}(2010)}]{johannsen-2010}
\bibinfo{author}{\bibfnamefont{T.}~\bibnamefont{Johannsen}} \bibnamefont{and}
  \bibinfo{author}{\bibfnamefont{D.}~\bibnamefont{Psaltis}},
  \bibinfo{journal}{Astrophys. J.} \textbf{\bibinfo{volume}{716}},
  \bibinfo{pages}{187} (\bibinfo{year}{2010}), \eprint{1003.3415}.

\bibitem[{\citenamefont{Johannsen and Psaltis}(2011)}]{johannsen-2011}
\bibinfo{author}{\bibfnamefont{T.}~\bibnamefont{Johannsen}} \bibnamefont{and}
  \bibinfo{author}{\bibfnamefont{D.}~\bibnamefont{Psaltis}},
  \bibinfo{journal}{Physical Review D} \textbf{\bibinfo{volume}{83}},
  \bibinfo{pages}{124015} (\bibinfo{year}{2011}).

\bibitem[{\citenamefont{Vigeland et~al.}(2011)\citenamefont{Vigeland, Yunes,
  and Stein}}]{vigeland-2011}
\bibinfo{author}{\bibfnamefont{S.}~\bibnamefont{Vigeland}},
  \bibinfo{author}{\bibfnamefont{N.}~\bibnamefont{Yunes}}, \bibnamefont{and}
  \bibinfo{author}{\bibfnamefont{L.}~\bibnamefont{Stein}},
  \bibinfo{journal}{Phys. Rev. D} \textbf{\bibinfo{volume}{83}},
  \bibinfo{pages}{104027} (\bibinfo{year}{2011}), \eprint{1102.3706}.

\bibitem[{\citenamefont{Johannsen}(2013)}]{Johannsen-2013}
\bibinfo{author}{\bibfnamefont{T.}~\bibnamefont{Johannsen}},
  \bibinfo{journal}{Phys. Rev. D} \textbf{\bibinfo{volume}{88}},
  \bibinfo{pages}{044002} (\bibinfo{year}{2013}), \eprint{1501.02809}.

\bibitem[{\citenamefont{Rezzolla and Zhidenko}(2014)}]{Rezzolla-2014}
\bibinfo{author}{\bibfnamefont{L.}~\bibnamefont{Rezzolla}} \bibnamefont{and}
  \bibinfo{author}{\bibfnamefont{A.}~\bibnamefont{Zhidenko}},
  \bibinfo{journal}{Phys. Rev. D} \textbf{\bibinfo{volume}{90}},
  \bibinfo{pages}{084009} (\bibinfo{year}{2014}), \eprint{1407.3086}.

\bibitem[{\citenamefont{Konoplya et~al.}(2016)\citenamefont{Konoplya, Rezzolla,
  and Zhidenko}}]{konoplya-2016b}
\bibinfo{author}{\bibfnamefont{R.}~\bibnamefont{Konoplya}},
  \bibinfo{author}{\bibfnamefont{L.}~\bibnamefont{Rezzolla}}, \bibnamefont{and}
  \bibinfo{author}{\bibfnamefont{A.}~\bibnamefont{Zhidenko}},
  \bibinfo{journal}{Phys. Rev. D} \textbf{\bibinfo{volume}{93}},
  \bibinfo{pages}{064015} (\bibinfo{year}{2016}), \eprint{1602.02378}.

\bibitem[{\citenamefont{Konoplya and Zhidenko}(2016)}]{konoplya_2016}
\bibinfo{author}{\bibfnamefont{R.}~\bibnamefont{Konoplya}} \bibnamefont{and}
  \bibinfo{author}{\bibfnamefont{A.}~\bibnamefont{Zhidenko}},
  \bibinfo{journal}{Physics Letters B} \textbf{\bibinfo{volume}{756}},
  \bibinfo{pages}{350} (\bibinfo{year}{2016}).

\bibitem[{\citenamefont{Ma and Rezzolla}(2024)}]{Ma-2024}
\bibinfo{author}{\bibfnamefont{Y.}~\bibnamefont{Ma}} \bibnamefont{and}
  \bibinfo{author}{\bibfnamefont{L.}~\bibnamefont{Rezzolla}}
  (\bibinfo{year}{2024}), \eprint{2404.06509}.

\bibitem[{\citenamefont{Ni et~al.}(2016)\citenamefont{Ni, Jiang, and
  Bambi}}]{Ni-2016}
\bibinfo{author}{\bibfnamefont{Y.}~\bibnamefont{Ni}},
  \bibinfo{author}{\bibfnamefont{J.}~\bibnamefont{Jiang}}, \bibnamefont{and}
  \bibinfo{author}{\bibfnamefont{C.}~\bibnamefont{Bambi}},
  \bibinfo{journal}{JCAP} \textbf{\bibinfo{volume}{09}}, \bibinfo{pages}{014}
  (\bibinfo{year}{2016}), \eprint{1607.04893}.

\bibitem[{\citenamefont{Cardenas-Avendano
  et~al.}(2020)\citenamefont{Cardenas-Avendano, Nampalliwar, and
  Yunes}}]{Cardenas-Avendano-2019}
\bibinfo{author}{\bibfnamefont{A.}~\bibnamefont{Cardenas-Avendano}},
  \bibinfo{author}{\bibfnamefont{S.}~\bibnamefont{Nampalliwar}},
  \bibnamefont{and} \bibinfo{author}{\bibfnamefont{N.}~\bibnamefont{Yunes}},
  \bibinfo{journal}{Class. Quant. Grav.} \textbf{\bibinfo{volume}{37}},
  \bibinfo{pages}{135008} (\bibinfo{year}{2020}), \eprint{1912.08062}.

\bibitem[{\citenamefont{Psaltis et~al.}(2020)}]{EHT-2020}
\bibinfo{author}{\bibfnamefont{D.}~\bibnamefont{Psaltis}} \bibnamefont{et~al.}
  (\bibinfo{collaboration}{Event Horizon Telescope}), \bibinfo{journal}{Phys.
  Rev. Lett.} \textbf{\bibinfo{volume}{125}}, \bibinfo{pages}{141104}
  (\bibinfo{year}{2020}), \eprint{2010.01055}.

\bibitem[{\citenamefont{Shashank and Bambi}(2022)}]{Shashank-2021}
\bibinfo{author}{\bibfnamefont{S.}~\bibnamefont{Shashank}} \bibnamefont{and}
  \bibinfo{author}{\bibfnamefont{C.}~\bibnamefont{Bambi}},
  \bibinfo{journal}{Phys. Rev. D} \textbf{\bibinfo{volume}{105}},
  \bibinfo{pages}{104004} (\bibinfo{year}{2022}), \eprint{2112.05388}.

\bibitem[{\citenamefont{Bambi}(2023)}]{Bambi-2023}
\bibinfo{author}{\bibfnamefont{C.}~\bibnamefont{Bambi}}, in
  \emph{\bibinfo{booktitle}{{35th International Workshop on High Energy
  Physics}: {From Quarks to Galaxies: Elucidating Dark Sides}}}
  (\bibinfo{year}{2023}), \eprint{2312.05857}.

\bibitem[{\citenamefont{Tripathi et~al.}(2024)\citenamefont{Tripathi, Mall, and
  Abdikamalov}}]{Tripathi-2024}
\bibinfo{author}{\bibfnamefont{A.}~\bibnamefont{Tripathi}},
  \bibinfo{author}{\bibfnamefont{G.}~\bibnamefont{Mall}}, \bibnamefont{and}
  \bibinfo{author}{\bibfnamefont{A.}~\bibnamefont{Abdikamalov}}
  (\bibinfo{year}{2024}), \eprint{2401.08545}.

\bibitem[{\citenamefont{Pringle}(1981)}]{Pringle-1981}
\bibinfo{author}{\bibfnamefont{J.~E.} \bibnamefont{Pringle}},
  \bibinfo{journal}{Annual review of astronomy and astrophysics}
  \textbf{\bibinfo{volume}{19}}, \bibinfo{pages}{137} (\bibinfo{year}{1981}).

\bibitem[{\citenamefont{Frank et~al.}(2002)\citenamefont{Frank, King, and
  Raine}}]{Frank-2002}
\bibinfo{author}{\bibfnamefont{J.}~\bibnamefont{Frank}},
  \bibinfo{author}{\bibfnamefont{A.}~\bibnamefont{King}}, \bibnamefont{and}
  \bibinfo{author}{\bibfnamefont{D.}~\bibnamefont{Raine}},
  \emph{\bibinfo{title}{Accretion power in astrophysics}}
  (\bibinfo{publisher}{Cambridge university press}, \bibinfo{year}{2002}).

\bibitem[{\citenamefont{Netzer}(2013)}]{netzer-2013}
\bibinfo{author}{\bibfnamefont{H.}~\bibnamefont{Netzer}},
  \emph{\bibinfo{title}{The physics and evolution of active galactic nuclei}}
  (\bibinfo{publisher}{Cambridge university press}, \bibinfo{year}{2013}).

\bibitem[{\citenamefont{Abramowicz and Fragile}(2013)}]{abramowicz-2013}
\bibinfo{author}{\bibfnamefont{M.~A.} \bibnamefont{Abramowicz}}
  \bibnamefont{and} \bibinfo{author}{\bibfnamefont{P.~C.}
  \bibnamefont{Fragile}}, \bibinfo{journal}{Living Reviews in Relativity}
  \textbf{\bibinfo{volume}{16}}, \bibinfo{pages}{1} (\bibinfo{year}{2013}).

\bibitem[{\citenamefont{Yuan and Narayan}(2014)}]{yuan-2014}
\bibinfo{author}{\bibfnamefont{F.}~\bibnamefont{Yuan}} \bibnamefont{and}
  \bibinfo{author}{\bibfnamefont{R.}~\bibnamefont{Narayan}},
  \bibinfo{journal}{Ann. Rev. Astron. Astrophys.}
  \textbf{\bibinfo{volume}{52}}, \bibinfo{pages}{529} (\bibinfo{year}{2014}),
  \eprint{1401.0586}.

\bibitem[{\citenamefont{Nandi et~al.}(2018)\citenamefont{Nandi, Mandal,
  Sreehari, Radhika, Das, Chattopadhyay, Iyer, Agrawal, and
  Aktar}}]{nandi-2018}
\bibinfo{author}{\bibfnamefont{A.}~\bibnamefont{Nandi}},
  \bibinfo{author}{\bibfnamefont{S.}~\bibnamefont{Mandal}},
  \bibinfo{author}{\bibfnamefont{H.}~\bibnamefont{Sreehari}},
  \bibinfo{author}{\bibfnamefont{D.}~\bibnamefont{Radhika}},
  \bibinfo{author}{\bibfnamefont{S.}~\bibnamefont{Das}},
  \bibinfo{author}{\bibfnamefont{I.}~\bibnamefont{Chattopadhyay}},
  \bibinfo{author}{\bibfnamefont{N.}~\bibnamefont{Iyer}},
  \bibinfo{author}{\bibfnamefont{V.}~\bibnamefont{Agrawal}}, \bibnamefont{and}
  \bibinfo{author}{\bibfnamefont{R.}~\bibnamefont{Aktar}},
  \bibinfo{journal}{Astrophysics and Space Science}
  \textbf{\bibinfo{volume}{363}}, \bibinfo{pages}{1} (\bibinfo{year}{2018}).

\bibitem[{\citenamefont{Das et~al.}(2021)\citenamefont{Das, Nandi, Agrawal,
  Dihingia, and Majumder}}]{das-2021}
\bibinfo{author}{\bibfnamefont{S.}~\bibnamefont{Das}},
  \bibinfo{author}{\bibfnamefont{A.}~\bibnamefont{Nandi}},
  \bibinfo{author}{\bibfnamefont{V.~K.} \bibnamefont{Agrawal}},
  \bibinfo{author}{\bibfnamefont{I.~K.} \bibnamefont{Dihingia}},
  \bibnamefont{and} \bibinfo{author}{\bibfnamefont{S.}~\bibnamefont{Majumder}},
  \bibinfo{journal}{Monthly Notices of the Royal Astronomical Society}
  \textbf{\bibinfo{volume}{507}}, \bibinfo{pages}{2777} (\bibinfo{year}{2021}).

\bibitem[{\citenamefont{Mondal et~al.}(2022{\natexlab{a}})\citenamefont{Mondal,
  Adhikari, Hryniewicz, Stalin, and Pandey}}]{mondal-2022a}
\bibinfo{author}{\bibfnamefont{S.}~\bibnamefont{Mondal}},
  \bibinfo{author}{\bibfnamefont{T.~P.} \bibnamefont{Adhikari}},
  \bibinfo{author}{\bibfnamefont{K.}~\bibnamefont{Hryniewicz}},
  \bibinfo{author}{\bibfnamefont{C.}~\bibnamefont{Stalin}}, \bibnamefont{and}
  \bibinfo{author}{\bibfnamefont{A.}~\bibnamefont{Pandey}},
  \bibinfo{journal}{Astronomy \& Astrophysics} \textbf{\bibinfo{volume}{662}},
  \bibinfo{pages}{A77} (\bibinfo{year}{2022}{\natexlab{a}}).

\bibitem[{\citenamefont{Mondal et~al.}(2022{\natexlab{b}})\citenamefont{Mondal,
  Palit, and Chakrabarti}}]{mondal-2022b}
\bibinfo{author}{\bibfnamefont{S.}~\bibnamefont{Mondal}},
  \bibinfo{author}{\bibfnamefont{B.}~\bibnamefont{Palit}}, \bibnamefont{and}
  \bibinfo{author}{\bibfnamefont{S.~K.} \bibnamefont{Chakrabarti}},
  \bibinfo{journal}{Journal of Astrophysics and Astronomy}
  \textbf{\bibinfo{volume}{43}}, \bibinfo{pages}{90}
  (\bibinfo{year}{2022}{\natexlab{b}}).

\bibitem[{\citenamefont{Palit and Mondal}(2023)}]{palit-2023}
\bibinfo{author}{\bibfnamefont{B.}~\bibnamefont{Palit}} \bibnamefont{and}
  \bibinfo{author}{\bibfnamefont{S.}~\bibnamefont{Mondal}},
  \bibinfo{journal}{Publications of the Astronomical Society of the Pacific}
  \textbf{\bibinfo{volume}{135}}, \bibinfo{pages}{054101}
  (\bibinfo{year}{2023}).

\bibitem[{\citenamefont{Mondal and Jithesh}(2023)}]{mondal-2023}
\bibinfo{author}{\bibfnamefont{S.}~\bibnamefont{Mondal}} \bibnamefont{and}
  \bibinfo{author}{\bibfnamefont{V.}~\bibnamefont{Jithesh}},
  \bibinfo{journal}{Monthly Notices of the Royal Astronomical Society}
  \textbf{\bibinfo{volume}{522}}, \bibinfo{pages}{2065} (\bibinfo{year}{2023}).

\bibitem[{\citenamefont{Heiland et~al.}(2023)\citenamefont{Heiland, Chatterjee,
  Safi-Harb, Jana, and Heyl}}]{heiland-2023}
\bibinfo{author}{\bibfnamefont{S.~R.} \bibnamefont{Heiland}},
  \bibinfo{author}{\bibfnamefont{A.}~\bibnamefont{Chatterjee}},
  \bibinfo{author}{\bibfnamefont{S.}~\bibnamefont{Safi-Harb}},
  \bibinfo{author}{\bibfnamefont{A.}~\bibnamefont{Jana}}, \bibnamefont{and}
  \bibinfo{author}{\bibfnamefont{J.}~\bibnamefont{Heyl}},
  \bibinfo{journal}{arXiv preprint arXiv:2307.06395}  (\bibinfo{year}{2023}).

\bibitem[{\citenamefont{Mondal et~al.}(2024)\citenamefont{Mondal, Suribhatla,
  Chatterjee, and Singh}}]{mondal-2024}
\bibinfo{author}{\bibfnamefont{S.}~\bibnamefont{Mondal}},
  \bibinfo{author}{\bibfnamefont{S.~P.} \bibnamefont{Suribhatla}},
  \bibinfo{author}{\bibfnamefont{K.}~\bibnamefont{Chatterjee}},
  \bibnamefont{and} \bibinfo{author}{\bibfnamefont{C.~B.} \bibnamefont{Singh}}
  (\bibinfo{year}{2024}), \eprint{2404.09643}.

\bibitem[{\citenamefont{Sreehari et~al.}(2020)\citenamefont{Sreehari, Nandi,
  Das, Agrawal, Mandal, Ramadevi, and Katoch}}]{sreehari-2020}
\bibinfo{author}{\bibfnamefont{H.}~\bibnamefont{Sreehari}},
  \bibinfo{author}{\bibfnamefont{A.}~\bibnamefont{Nandi}},
  \bibinfo{author}{\bibfnamefont{S.}~\bibnamefont{Das}},
  \bibinfo{author}{\bibfnamefont{V.}~\bibnamefont{Agrawal}},
  \bibinfo{author}{\bibfnamefont{S.}~\bibnamefont{Mandal}},
  \bibinfo{author}{\bibfnamefont{M.}~\bibnamefont{Ramadevi}}, \bibnamefont{and}
  \bibinfo{author}{\bibfnamefont{T.}~\bibnamefont{Katoch}},
  \bibinfo{journal}{Monthly Notices of the Royal Astronomical Society}
  \textbf{\bibinfo{volume}{499}}, \bibinfo{pages}{5891} (\bibinfo{year}{2020}).

\bibitem[{\citenamefont{Sriram et~al.}(2021)\citenamefont{Sriram, Harikrishna,
  and Choi}}]{Sriram-2021}
\bibinfo{author}{\bibfnamefont{K.}~\bibnamefont{Sriram}},
  \bibinfo{author}{\bibfnamefont{S.}~\bibnamefont{Harikrishna}},
  \bibnamefont{and} \bibinfo{author}{\bibfnamefont{C.~S.} \bibnamefont{Choi}},
  \bibinfo{journal}{Astrophys. J.} \textbf{\bibinfo{volume}{911}},
  \bibinfo{pages}{127} (\bibinfo{year}{2021}), \eprint{2103.02422}.

\bibitem[{\citenamefont{Majumder et~al.}(2022)\citenamefont{Majumder, Sreehari,
  Aftab, Katoch, Das, and Nandi}}]{majumder-2022}
\bibinfo{author}{\bibfnamefont{S.}~\bibnamefont{Majumder}},
  \bibinfo{author}{\bibfnamefont{H.}~\bibnamefont{Sreehari}},
  \bibinfo{author}{\bibfnamefont{N.}~\bibnamefont{Aftab}},
  \bibinfo{author}{\bibfnamefont{T.}~\bibnamefont{Katoch}},
  \bibinfo{author}{\bibfnamefont{S.}~\bibnamefont{Das}}, \bibnamefont{and}
  \bibinfo{author}{\bibfnamefont{A.}~\bibnamefont{Nandi}},
  \bibinfo{journal}{Monthly Notices of the Royal Astronomical Society}
  \textbf{\bibinfo{volume}{512}}, \bibinfo{pages}{2508} (\bibinfo{year}{2022}).

\bibitem[{\citenamefont{Dhaka et~al.}(2023)\citenamefont{Dhaka, Misra, Yadav,
  and Jain}}]{dhaka-2023}
\bibinfo{author}{\bibfnamefont{R.}~\bibnamefont{Dhaka}},
  \bibinfo{author}{\bibfnamefont{R.}~\bibnamefont{Misra}},
  \bibinfo{author}{\bibfnamefont{J.}~\bibnamefont{Yadav}}, \bibnamefont{and}
  \bibinfo{author}{\bibfnamefont{P.}~\bibnamefont{Jain}},
  \bibinfo{journal}{Monthly Notices of the Royal Astronomical Society} p.
  \bibinfo{pages}{stad2075} (\bibinfo{year}{2023}).

\bibitem[{\citenamefont{Rawat et~al.}(2023{\natexlab{a}})\citenamefont{Rawat,
  M\'endez, Garc\'\i{}a, Altamirano, Karpouzas, Zhang, Alabarta, Belloni, Jain,
  and Bellavita}}]{Rawat-2023a}
\bibinfo{author}{\bibfnamefont{D.}~\bibnamefont{Rawat}},
  \bibinfo{author}{\bibfnamefont{M.}~\bibnamefont{M\'endez}},
  \bibinfo{author}{\bibfnamefont{F.}~\bibnamefont{Garc\'\i{}a}},
  \bibinfo{author}{\bibfnamefont{D.}~\bibnamefont{Altamirano}},
  \bibinfo{author}{\bibfnamefont{K.}~\bibnamefont{Karpouzas}},
  \bibinfo{author}{\bibfnamefont{L.}~\bibnamefont{Zhang}},
  \bibinfo{author}{\bibfnamefont{K.}~\bibnamefont{Alabarta}},
  \bibinfo{author}{\bibfnamefont{T.~M.} \bibnamefont{Belloni}},
  \bibinfo{author}{\bibfnamefont{P.}~\bibnamefont{Jain}}, \bibnamefont{and}
  \bibinfo{author}{\bibfnamefont{C.}~\bibnamefont{Bellavita}},
  \bibinfo{journal}{Mon. Not. Roy. Astron. Soc.}
  \textbf{\bibinfo{volume}{520}}, \bibinfo{pages}{113}
  (\bibinfo{year}{2023}{\natexlab{a}}), \eprint{2301.04418}.

\bibitem[{\citenamefont{Rawat et~al.}(2023{\natexlab{b}})\citenamefont{Rawat,
  Husain, and Misra}}]{Rawat-2023c}
\bibinfo{author}{\bibfnamefont{D.}~\bibnamefont{Rawat}},
  \bibinfo{author}{\bibfnamefont{N.}~\bibnamefont{Husain}}, \bibnamefont{and}
  \bibinfo{author}{\bibfnamefont{R.}~\bibnamefont{Misra}}
  (\bibinfo{year}{2023}{\natexlab{b}}), \eprint{2307.11460}.

\bibitem[{\citenamefont{Patra et~al.}(2022)\citenamefont{Patra, Majhi, and
  Das}}]{Patra-2022}
\bibinfo{author}{\bibfnamefont{S.}~\bibnamefont{Patra}},
  \bibinfo{author}{\bibfnamefont{B.~R.} \bibnamefont{Majhi}}, \bibnamefont{and}
  \bibinfo{author}{\bibfnamefont{S.}~\bibnamefont{Das}},
  \bibinfo{journal}{Phys. Dark Univ.} \textbf{\bibinfo{volume}{37}},
  \bibinfo{pages}{101120} (\bibinfo{year}{2022}), \eprint{2202.10863}.

\bibitem[{\citenamefont{Patra et~al.}(2024)\citenamefont{Patra, Majhi, and
  Das}}]{Patra-2023}
\bibinfo{author}{\bibfnamefont{S.}~\bibnamefont{Patra}},
  \bibinfo{author}{\bibfnamefont{B.~R.} \bibnamefont{Majhi}}, \bibnamefont{and}
  \bibinfo{author}{\bibfnamefont{S.}~\bibnamefont{Das}},
  \bibinfo{journal}{JCAP} \textbf{\bibinfo{volume}{01}}, \bibinfo{pages}{060}
  (\bibinfo{year}{2024}), \eprint{2308.12839}.

\bibitem[{\citenamefont{Abbott et~al.}(2016{\natexlab{a}})\citenamefont{Abbott,
  Abbott, Abbott, Abernathy, Acernese, Ackley, Adams, Adams, Addesso, Adhikari
  et~al.}}]{abbott-2016a}
\bibinfo{author}{\bibfnamefont{B.~P.} \bibnamefont{Abbott}},
  \bibinfo{author}{\bibfnamefont{R.}~\bibnamefont{Abbott}},
  \bibinfo{author}{\bibfnamefont{T.}~\bibnamefont{Abbott}},
  \bibinfo{author}{\bibfnamefont{M.}~\bibnamefont{Abernathy}},
  \bibinfo{author}{\bibfnamefont{F.}~\bibnamefont{Acernese}},
  \bibinfo{author}{\bibfnamefont{K.}~\bibnamefont{Ackley}},
  \bibinfo{author}{\bibfnamefont{C.}~\bibnamefont{Adams}},
  \bibinfo{author}{\bibfnamefont{T.}~\bibnamefont{Adams}},
  \bibinfo{author}{\bibfnamefont{P.}~\bibnamefont{Addesso}},
  \bibinfo{author}{\bibfnamefont{R.}~\bibnamefont{Adhikari}},
  \bibnamefont{et~al.}, \bibinfo{journal}{Physical review letters}
  \textbf{\bibinfo{volume}{116}}, \bibinfo{pages}{061102}
  (\bibinfo{year}{2016}{\natexlab{a}}).

\bibitem[{\citenamefont{Abbott et~al.}(2016{\natexlab{b}})}]{abbott-2016-b}
\bibinfo{author}{\bibfnamefont{B.~P.} \bibnamefont{Abbott}}
  \bibnamefont{et~al.} (\bibinfo{collaboration}{LIGO Scientific, Virgo}),
  \bibinfo{journal}{Phys. Rev. Lett.} \textbf{\bibinfo{volume}{116}},
  \bibinfo{pages}{221101} (\bibinfo{year}{2016}{\natexlab{b}}),
  \bibinfo{note}{[Erratum: Phys.Rev.Lett. 121, 129902 (2018)]},
  \eprint{1602.03841}.

\bibitem[{\citenamefont{Barman and Mukherjee}(2021)}]{Barman-2021}
\bibinfo{author}{\bibfnamefont{S.}~\bibnamefont{Barman}} \bibnamefont{and}
  \bibinfo{author}{\bibfnamefont{S.}~\bibnamefont{Mukherjee}},
  \bibinfo{journal}{Eur. Phys. J. C} \textbf{\bibinfo{volume}{81}},
  \bibinfo{pages}{453} (\bibinfo{year}{2021}), \eprint{2102.04066}.

\bibitem[{\citenamefont{Rezzolla and Zanotti}(2013)}]{Rezzolla-2013}
\bibinfo{author}{\bibfnamefont{L.}~\bibnamefont{Rezzolla}} \bibnamefont{and}
  \bibinfo{author}{\bibfnamefont{O.}~\bibnamefont{Zanotti}},
  \emph{\bibinfo{title}{{Relativistic Hydrodynamics}}}
  (\bibinfo{publisher}{Oxford University Press}, \bibinfo{year}{2013}), ISBN
  \bibinfo{isbn}{978-0-19-174650-5, 978-0-19-852890-6}.

\bibitem[{\citenamefont{Taub}(1948)}]{Taub-1948}
\bibinfo{author}{\bibfnamefont{A.~H.} \bibnamefont{Taub}},
  \bibinfo{journal}{Phys. Rev.} \textbf{\bibinfo{volume}{74}},
  \bibinfo{pages}{328} (\bibinfo{year}{1948}).

\bibitem[{\citenamefont{Novikov and Thorne}(1973)}]{novikov-1973}
\bibinfo{author}{\bibfnamefont{I.~D.} \bibnamefont{Novikov}} \bibnamefont{and}
  \bibinfo{author}{\bibfnamefont{K.~S.} \bibnamefont{Thorne}},
  \bibinfo{journal}{Black holes (Les astres occlus)}
  \textbf{\bibinfo{volume}{1}}, \bibinfo{pages}{343} (\bibinfo{year}{1973}).

\bibitem[{\citenamefont{Dihingia et~al.}(2019)\citenamefont{Dihingia, Das,
  Maity, and Nandi}}]{Dihingia-2019}
\bibinfo{author}{\bibfnamefont{I.}~\bibnamefont{Dihingia}},
  \bibinfo{author}{\bibfnamefont{S.}~\bibnamefont{Das}},
  \bibinfo{author}{\bibfnamefont{D.}~\bibnamefont{Maity}}, \bibnamefont{and}
  \bibinfo{author}{\bibfnamefont{A.}~\bibnamefont{Nandi}},
  \bibinfo{journal}{Mon. Not. Roy. Astron. Soc.}
  \textbf{\bibinfo{volume}{488}}, \bibinfo{pages}{2412} (\bibinfo{year}{2019}),
  \eprint{1903.02856}.

\bibitem[{\citenamefont{Boyer and Lindquist}(1967)}]{Boyer_1966}
\bibinfo{author}{\bibfnamefont{R.~H.} \bibnamefont{Boyer}} \bibnamefont{and}
  \bibinfo{author}{\bibfnamefont{R.~W.} \bibnamefont{Lindquist}},
  \bibinfo{journal}{J. Math. Phys.} \textbf{\bibinfo{volume}{8}},
  \bibinfo{pages}{265} (\bibinfo{year}{1967}).

\bibitem[{\citenamefont{Landau and Lifshitz}(2013)}]{landau-2013}
\bibinfo{author}{\bibfnamefont{L.~D.} \bibnamefont{Landau}} \bibnamefont{and}
  \bibinfo{author}{\bibfnamefont{E.~M.} \bibnamefont{Lifshitz}},
  \emph{\bibinfo{title}{Fluid mechanics: Landau And Lifshitz: course of
  theoretical physics, Volume 6}}, vol.~\bibinfo{volume}{6}
  (\bibinfo{publisher}{Elsevier}, \bibinfo{year}{2013}).

\bibitem[{\citenamefont{Peitz and Appl}(1997)}]{Peitz-1996}
\bibinfo{author}{\bibfnamefont{J.}~\bibnamefont{Peitz}} \bibnamefont{and}
  \bibinfo{author}{\bibfnamefont{S.}~\bibnamefont{Appl}},
  \bibinfo{journal}{Mon. Not. Roy. Astron. Soc.}
  \textbf{\bibinfo{volume}{286}}, \bibinfo{pages}{681} (\bibinfo{year}{1997}),
  \eprint{astro-ph/9612205}.

\bibitem[{\citenamefont{Chattopadhyay and Kumar}(2016)}]{Chattopadhyay_2016}
\bibinfo{author}{\bibfnamefont{I.}~\bibnamefont{Chattopadhyay}}
  \bibnamefont{and} \bibinfo{author}{\bibfnamefont{R.}~\bibnamefont{Kumar}},
  \bibinfo{journal}{Mon. Not. Roy. Astron. Soc.}
  \textbf{\bibinfo{volume}{459}}, \bibinfo{pages}{3792} (\bibinfo{year}{2016}),
  \eprint{1605.00752}.

\bibitem[{\citenamefont{Gammie and Popham}(1998)}]{Gammie_1997}
\bibinfo{author}{\bibfnamefont{C.~F.} \bibnamefont{Gammie}} \bibnamefont{and}
  \bibinfo{author}{\bibfnamefont{R.}~\bibnamefont{Popham}},
  \bibinfo{journal}{Astrophys. J.} \textbf{\bibinfo{volume}{498}},
  \bibinfo{pages}{313} (\bibinfo{year}{1998}), \eprint{astro-ph/9705117}.

\bibitem[{\citenamefont{Popham and Gammie}(1998)}]{Popham_1998}
\bibinfo{author}{\bibfnamefont{R.}~\bibnamefont{Popham}} \bibnamefont{and}
  \bibinfo{author}{\bibfnamefont{C.~F.} \bibnamefont{Gammie}},
  \bibinfo{journal}{Astrophys. J.} \textbf{\bibinfo{volume}{504}},
  \bibinfo{pages}{419} (\bibinfo{year}{1998}), \eprint{astro-ph/9802321}.

\bibitem[{\citenamefont{Lasota}(1994)}]{lasota-1994}
\bibinfo{author}{\bibfnamefont{J.}~\bibnamefont{Lasota}}, in
  \emph{\bibinfo{booktitle}{Theory of Accretion Disks—2: Proceedings of the
  NATO Advanced Research Workshop on Theory of Accretion Disks—2 Garching,
  Germany March 22--26, 1993}} (\bibinfo{organization}{Springer},
  \bibinfo{year}{1994}), pp. \bibinfo{pages}{341--349}.

\bibitem[{\citenamefont{Riffert and Herold}(1995)}]{riffert-1995}
\bibinfo{author}{\bibfnamefont{H.}~\bibnamefont{Riffert}} \bibnamefont{and}
  \bibinfo{author}{\bibfnamefont{H.}~\bibnamefont{Herold}},
  \bibinfo{journal}{Astrophysical Journal v. 450, p. 508}
  \textbf{\bibinfo{volume}{450}}, \bibinfo{pages}{508} (\bibinfo{year}{1995}).

\bibitem[{\citenamefont{Lu}(1985)}]{lu-1985}
\bibinfo{author}{\bibfnamefont{J.}~\bibnamefont{Lu}},
  \bibinfo{journal}{Astronomy and Astrophysics} \textbf{\bibinfo{volume}{148}},
  \bibinfo{pages}{176} (\bibinfo{year}{1985}).

\bibitem[{\citenamefont{Chattopadhyay and Ryu}(2009)}]{chattopadhyay-2009}
\bibinfo{author}{\bibfnamefont{I.}~\bibnamefont{Chattopadhyay}}
  \bibnamefont{and} \bibinfo{author}{\bibfnamefont{D.}~\bibnamefont{Ryu}},
  \bibinfo{journal}{The Astrophysical Journal} \textbf{\bibinfo{volume}{694}},
  \bibinfo{pages}{492} (\bibinfo{year}{2009}).

\bibitem[{\citenamefont{Dihingia
  et~al.}(2020{\natexlab{a}})\citenamefont{Dihingia, Das, Prabhakar, and
  Mandal}}]{dihingia-2020}
\bibinfo{author}{\bibfnamefont{I.~K.} \bibnamefont{Dihingia}},
  \bibinfo{author}{\bibfnamefont{S.}~\bibnamefont{Das}},
  \bibinfo{author}{\bibfnamefont{G.}~\bibnamefont{Prabhakar}},
  \bibnamefont{and} \bibinfo{author}{\bibfnamefont{S.}~\bibnamefont{Mandal}},
  \bibinfo{journal}{Monthly Notices of the Royal Astronomical Society}
  \textbf{\bibinfo{volume}{496}}, \bibinfo{pages}{3043}
  (\bibinfo{year}{2020}{\natexlab{a}}).

\bibitem[{\citenamefont{Sarkar and Chattopadhyay}(2022)}]{Sarkar-2022}
\bibinfo{author}{\bibfnamefont{S.}~\bibnamefont{Sarkar}} \bibnamefont{and}
  \bibinfo{author}{\bibfnamefont{I.}~\bibnamefont{Chattopadhyay}},
  \bibinfo{journal}{Journal of Astrophysics and Astronomy}
  \textbf{\bibinfo{volume}{43}}, \bibinfo{pages}{34} (\bibinfo{year}{2022}).

\bibitem[{\citenamefont{Svensson}(1982)}]{Svensson-1982}
\bibinfo{author}{\bibfnamefont{R.}~\bibnamefont{Svensson}},
  \bibinfo{journal}{Astrophys. J.} \textbf{\bibinfo{volume}{258}},
  \bibinfo{pages}{335} (\bibinfo{year}{1982}).

\bibitem[{\citenamefont{Nozawa et~al.}(2009)\citenamefont{Nozawa, Takahashi,
  Kohyama, and Itoh}}]{nozawa-2009}
\bibinfo{author}{\bibfnamefont{S.}~\bibnamefont{Nozawa}},
  \bibinfo{author}{\bibfnamefont{K.}~\bibnamefont{Takahashi}},
  \bibinfo{author}{\bibfnamefont{Y.}~\bibnamefont{Kohyama}}, \bibnamefont{and}
  \bibinfo{author}{\bibfnamefont{N.}~\bibnamefont{Itoh}},
  \bibinfo{journal}{Astronomy \& Astrophysics} \textbf{\bibinfo{volume}{499}},
  \bibinfo{pages}{661} (\bibinfo{year}{2009}).

\bibitem[{\citenamefont{Yarza et~al.}(2020)\citenamefont{Yarza, Wong, Ryan, and
  Gammie}}]{Yarza-2020}
\bibinfo{author}{\bibfnamefont{R.}~\bibnamefont{Yarza}},
  \bibinfo{author}{\bibfnamefont{G.~N.} \bibnamefont{Wong}},
  \bibinfo{author}{\bibfnamefont{B.~R.} \bibnamefont{Ryan}}, \bibnamefont{and}
  \bibinfo{author}{\bibfnamefont{C.~F.} \bibnamefont{Gammie}},
  \bibinfo{journal}{Astrophys. J.} \textbf{\bibinfo{volume}{898}},
  \bibinfo{pages}{50} (\bibinfo{year}{2020}), \eprint{2006.01145}.

\bibitem[{\citenamefont{Chattopadhyay and
  Chakrabarti}(2000)}]{chattopadhyay-2000}
\bibinfo{author}{\bibfnamefont{I.}~\bibnamefont{Chattopadhyay}}
  \bibnamefont{and} \bibinfo{author}{\bibfnamefont{S.~K.}
  \bibnamefont{Chakrabarti}}, \bibinfo{journal}{International Journal of Modern
  Physics D} \textbf{\bibinfo{volume}{9}}, \bibinfo{pages}{717}
  (\bibinfo{year}{2000}).

\bibitem[{\citenamefont{Dihingia
  et~al.}(2020{\natexlab{b}})\citenamefont{Dihingia, Maity, Chakrabarti, and
  Das}}]{dihingia-2020a}
\bibinfo{author}{\bibfnamefont{I.~K.} \bibnamefont{Dihingia}},
  \bibinfo{author}{\bibfnamefont{D.}~\bibnamefont{Maity}},
  \bibinfo{author}{\bibfnamefont{S.}~\bibnamefont{Chakrabarti}},
  \bibnamefont{and} \bibinfo{author}{\bibfnamefont{S.}~\bibnamefont{Das}},
  \bibinfo{journal}{Physical Review D} \textbf{\bibinfo{volume}{102}},
  \bibinfo{pages}{023012} (\bibinfo{year}{2020}{\natexlab{b}}).

\bibitem[{\citenamefont{Sen et~al.}(2022)\citenamefont{Sen, Maity, and
  Das}}]{Sen-2022}
\bibinfo{author}{\bibfnamefont{G.}~\bibnamefont{Sen}},
  \bibinfo{author}{\bibfnamefont{D.}~\bibnamefont{Maity}}, \bibnamefont{and}
  \bibinfo{author}{\bibfnamefont{S.}~\bibnamefont{Das}},
  \bibinfo{journal}{JCAP} \textbf{\bibinfo{volume}{08}}, \bibinfo{pages}{048}
  (\bibinfo{year}{2022}), \eprint{2204.02110}.

\bibitem[{\citenamefont{Liang and Thompson}(1980)}]{liang-1980}
\bibinfo{author}{\bibfnamefont{E.}~\bibnamefont{Liang}} \bibnamefont{and}
  \bibinfo{author}{\bibfnamefont{K.}~\bibnamefont{Thompson}},
  \bibinfo{journal}{The Astrophysical Journal} \textbf{\bibinfo{volume}{240}},
  \bibinfo{pages}{271} (\bibinfo{year}{1980}).

\bibitem[{\citenamefont{Abramowicz and Zurek}(1981)}]{abramowicz-1981}
\bibinfo{author}{\bibfnamefont{M.~A.} \bibnamefont{Abramowicz}}
  \bibnamefont{and} \bibinfo{author}{\bibfnamefont{W.}~\bibnamefont{Zurek}},
  \bibinfo{journal}{The Astrophysical Journal} \textbf{\bibinfo{volume}{246}},
  \bibinfo{pages}{314} (\bibinfo{year}{1981}).

\bibitem[{\citenamefont{{Chakrabarti}}(1996)}]{Chakrabarti-1996}
\bibinfo{author}{\bibfnamefont{S.~K.} \bibnamefont{{Chakrabarti}}},
  \bibinfo{journal}{\apj} \textbf{\bibinfo{volume}{464}}, \bibinfo{pages}{664}
  (\bibinfo{year}{1996}), \eprint{astro-ph/9606145}.

\bibitem[{\citenamefont{Dihingia et~al.}(2018)\citenamefont{Dihingia, Das,
  Maity, and Chakrabarti}}]{dihingia-2018}
\bibinfo{author}{\bibfnamefont{I.~K.} \bibnamefont{Dihingia}},
  \bibinfo{author}{\bibfnamefont{S.}~\bibnamefont{Das}},
  \bibinfo{author}{\bibfnamefont{D.}~\bibnamefont{Maity}}, \bibnamefont{and}
  \bibinfo{author}{\bibfnamefont{S.}~\bibnamefont{Chakrabarti}},
  \bibinfo{journal}{Physical Review D} \textbf{\bibinfo{volume}{98}},
  \bibinfo{pages}{083004} (\bibinfo{year}{2018}).

\bibitem[{\citenamefont{Mitra and Das}(2024)}]{Mitra-2024}
\bibinfo{author}{\bibfnamefont{S.}~\bibnamefont{Mitra}} \bibnamefont{and}
  \bibinfo{author}{\bibfnamefont{S.}~\bibnamefont{Das}} (\bibinfo{year}{2024}),
  \eprint{2405.16326}.

\bibitem[{\citenamefont{Becker and Kazanas}(2001)}]{Becker-2001}
\bibinfo{author}{\bibfnamefont{P.~A.} \bibnamefont{Becker}} \bibnamefont{and}
  \bibinfo{author}{\bibfnamefont{D.}~\bibnamefont{Kazanas}},
  \bibinfo{journal}{Astrophys. J.} \textbf{\bibinfo{volume}{546}},
  \bibinfo{pages}{429} (\bibinfo{year}{2001}), \eprint{astro-ph/0101020}.

\bibitem[{\citenamefont{Nandi et~al.}(2024)\citenamefont{Nandi, Das, Majumder,
  Katoch, Antia, and Shah}}]{nandi-2024}
\bibinfo{author}{\bibfnamefont{A.}~\bibnamefont{Nandi}},
  \bibinfo{author}{\bibfnamefont{S.}~\bibnamefont{Das}},
  \bibinfo{author}{\bibfnamefont{S.}~\bibnamefont{Majumder}},
  \bibinfo{author}{\bibfnamefont{T.}~\bibnamefont{Katoch}},
  \bibinfo{author}{\bibfnamefont{H.~M.} \bibnamefont{Antia}}, \bibnamefont{and}
  \bibinfo{author}{\bibfnamefont{P.}~\bibnamefont{Shah}}
  (\bibinfo{year}{2024}), \eprint{2404.17160}.

\bibitem[{\citenamefont{Chatterjee et~al.}(2024)\citenamefont{Chatterjee,
  Mondal, Singh, and Sugizaki}}]{Chatterjee-2024}
\bibinfo{author}{\bibfnamefont{K.}~\bibnamefont{Chatterjee}},
  \bibinfo{author}{\bibfnamefont{S.}~\bibnamefont{Mondal}},
  \bibinfo{author}{\bibfnamefont{C.~B.} \bibnamefont{Singh}}, \bibnamefont{and}
  \bibinfo{author}{\bibfnamefont{M.}~\bibnamefont{Sugizaki}}
  (\bibinfo{year}{2024}), \eprint{2405.01498}.

\bibitem[{\citenamefont{Molteni et~al.}(1996)\citenamefont{Molteni, Sponholz,
  and Chakrabarti}}]{molteni-1995}
\bibinfo{author}{\bibfnamefont{D.~M.} \bibnamefont{Molteni}},
  \bibinfo{author}{\bibfnamefont{H.}~\bibnamefont{Sponholz}}, \bibnamefont{and}
  \bibinfo{author}{\bibfnamefont{S.~K.} \bibnamefont{Chakrabarti}},
  \bibinfo{journal}{Astrophys. J.} \textbf{\bibinfo{volume}{457}},
  \bibinfo{pages}{805} (\bibinfo{year}{1996}), \eprint{astro-ph/9508022}.

\bibitem[{\citenamefont{Chakrabarti et~al.}(2015)\citenamefont{Chakrabarti,
  Mondal, and Debnath}}]{Chakrabarti-2015}
\bibinfo{author}{\bibfnamefont{S.~K.} \bibnamefont{Chakrabarti}},
  \bibinfo{author}{\bibfnamefont{S.}~\bibnamefont{Mondal}}, \bibnamefont{and}
  \bibinfo{author}{\bibfnamefont{D.}~\bibnamefont{Debnath}},
  \bibinfo{journal}{Mon. Not. Roy. Astron. Soc.}
  \textbf{\bibinfo{volume}{452}}, \bibinfo{pages}{3451} (\bibinfo{year}{2015}),
  \eprint{1507.02831}.

\bibitem[{\citenamefont{Kim et~al.}(2019)\citenamefont{Kim, Garain,
  Chakrabarti, and Balsara}}]{Kim-2018}
\bibinfo{author}{\bibfnamefont{J.}~\bibnamefont{Kim}},
  \bibinfo{author}{\bibfnamefont{S.~K.} \bibnamefont{Garain}},
  \bibinfo{author}{\bibfnamefont{S.~K.} \bibnamefont{Chakrabarti}},
  \bibnamefont{and} \bibinfo{author}{\bibfnamefont{D.~S.}
  \bibnamefont{Balsara}}, \bibinfo{journal}{Mon. Not. Roy. Astron. Soc.}
  \textbf{\bibinfo{volume}{482}}, \bibinfo{pages}{3636} (\bibinfo{year}{2019}),
  \eprint{1810.12469}.

\bibitem[{\citenamefont{Remillard and McClintock}(2006)}]{Remillard-2006a}
\bibinfo{author}{\bibfnamefont{R.~A.} \bibnamefont{Remillard}}
  \bibnamefont{and} \bibinfo{author}{\bibfnamefont{J.~E.}
  \bibnamefont{McClintock}}, \bibinfo{journal}{Annu. Rev. Astron. Astrophys.}
  \textbf{\bibinfo{volume}{44}}, \bibinfo{pages}{49} (\bibinfo{year}{2006}).

\bibitem[{\citenamefont{Homan et~al.}(2001)\citenamefont{Homan, Wijnands,
  van~der Klis, Belloni, van Paradijs, Klein-Wolt, Fender, and
  Mendez}}]{Homan-2000}
\bibinfo{author}{\bibfnamefont{J.}~\bibnamefont{Homan}},
  \bibinfo{author}{\bibfnamefont{R.}~\bibnamefont{Wijnands}},
  \bibinfo{author}{\bibfnamefont{M.}~\bibnamefont{van~der Klis}},
  \bibinfo{author}{\bibfnamefont{T.}~\bibnamefont{Belloni}},
  \bibinfo{author}{\bibfnamefont{J.}~\bibnamefont{van Paradijs}},
  \bibinfo{author}{\bibfnamefont{M.}~\bibnamefont{Klein-Wolt}},
  \bibinfo{author}{\bibfnamefont{R.~P.} \bibnamefont{Fender}},
  \bibnamefont{and} \bibinfo{author}{\bibfnamefont{M.}~\bibnamefont{Mendez}},
  \bibinfo{journal}{Astrophys. J. Suppl.} \textbf{\bibinfo{volume}{132}},
  \bibinfo{pages}{377} (\bibinfo{year}{2001}), \eprint{astro-ph/0001163}.

\bibitem[{\citenamefont{Miller et~al.}(2001)\citenamefont{Miller, Wijnands,
  Homan, Belloni, Pooley, Corbel, Kouveliotou, van~der Klis, and
  Lewin}}]{Miller-2001}
\bibinfo{author}{\bibfnamefont{J.~M.} \bibnamefont{Miller}},
  \bibinfo{author}{\bibfnamefont{R.}~\bibnamefont{Wijnands}},
  \bibinfo{author}{\bibfnamefont{J.}~\bibnamefont{Homan}},
  \bibinfo{author}{\bibfnamefont{T.}~\bibnamefont{Belloni}},
  \bibinfo{author}{\bibfnamefont{D.}~\bibnamefont{Pooley}},
  \bibinfo{author}{\bibfnamefont{S.}~\bibnamefont{Corbel}},
  \bibinfo{author}{\bibfnamefont{C.}~\bibnamefont{Kouveliotou}},
  \bibinfo{author}{\bibfnamefont{M.}~\bibnamefont{van~der Klis}},
  \bibnamefont{and} \bibinfo{author}{\bibfnamefont{W.~H.~G.}
  \bibnamefont{Lewin}}, \bibinfo{journal}{Astrophys. J.}
  \textbf{\bibinfo{volume}{563}}, \bibinfo{pages}{928} (\bibinfo{year}{2001}),
  \eprint{astro-ph/0105371}.

\bibitem[{\citenamefont{Remillard et~al.}(2002)\citenamefont{Remillard, Muno,
  McClintock, and Orosz}}]{Remillard-2002}
\bibinfo{author}{\bibfnamefont{R.~A.} \bibnamefont{Remillard}},
  \bibinfo{author}{\bibfnamefont{M.~P.} \bibnamefont{Muno}},
  \bibinfo{author}{\bibfnamefont{J.~E.} \bibnamefont{McClintock}},
  \bibnamefont{and} \bibinfo{author}{\bibfnamefont{J.~A.} \bibnamefont{Orosz}},
  \bibinfo{journal}{Astrophys. J.} \textbf{\bibinfo{volume}{580}},
  \bibinfo{pages}{1030} (\bibinfo{year}{2002}), \eprint{astro-ph/0202305}.

\bibitem[{\citenamefont{Homan et~al.}(2005)\citenamefont{Homan, Miller,
  Wijnands, van~der Klis, Belloni, Steeghs, and Lewin}}]{Homan-2005}
\bibinfo{author}{\bibfnamefont{J.}~\bibnamefont{Homan}},
  \bibinfo{author}{\bibfnamefont{J.~M.} \bibnamefont{Miller}},
  \bibinfo{author}{\bibfnamefont{R.}~\bibnamefont{Wijnands}},
  \bibinfo{author}{\bibfnamefont{M.}~\bibnamefont{van~der Klis}},
  \bibinfo{author}{\bibfnamefont{T.}~\bibnamefont{Belloni}},
  \bibinfo{author}{\bibfnamefont{D.}~\bibnamefont{Steeghs}}, \bibnamefont{and}
  \bibinfo{author}{\bibfnamefont{W.~H.~G.} \bibnamefont{Lewin}},
  \bibinfo{journal}{Astrophys. J.} \textbf{\bibinfo{volume}{623}},
  \bibinfo{pages}{383} (\bibinfo{year}{2005}), \eprint{astro-ph/0406334}.

\bibitem[{\citenamefont{Remillard et~al.}(2006)\citenamefont{Remillard,
  McClintock, Orosz, and Levine}}]{Remillard-2006b}
\bibinfo{author}{\bibfnamefont{R.~A.} \bibnamefont{Remillard}},
  \bibinfo{author}{\bibfnamefont{J.~E.} \bibnamefont{McClintock}},
  \bibinfo{author}{\bibfnamefont{J.~A.} \bibnamefont{Orosz}}, \bibnamefont{and}
  \bibinfo{author}{\bibfnamefont{A.~M.} \bibnamefont{Levine}},
  \bibinfo{journal}{Astrophys. J.} \textbf{\bibinfo{volume}{637}},
  \bibinfo{pages}{1002} (\bibinfo{year}{2006}), \eprint{astro-ph/0407025}.

\bibitem[{\citenamefont{Orosz et~al.}(2011)\citenamefont{Orosz, Steiner,
  McClintock, Torres, Remillard, Bailyn, and Miller}}]{Orosz-2011}
\bibinfo{author}{\bibfnamefont{J.~A.} \bibnamefont{Orosz}},
  \bibinfo{author}{\bibfnamefont{J.~F.} \bibnamefont{Steiner}},
  \bibinfo{author}{\bibfnamefont{J.~E.} \bibnamefont{McClintock}},
  \bibinfo{author}{\bibfnamefont{M.~A.~P.} \bibnamefont{Torres}},
  \bibinfo{author}{\bibfnamefont{R.~A.} \bibnamefont{Remillard}},
  \bibinfo{author}{\bibfnamefont{C.~D.} \bibnamefont{Bailyn}},
  \bibnamefont{and} \bibinfo{author}{\bibfnamefont{J.~M.}
  \bibnamefont{Miller}}, \bibinfo{journal}{Astrophys. J.}
  \textbf{\bibinfo{volume}{730}}, \bibinfo{pages}{75} (\bibinfo{year}{2011}),
  \eprint{1101.2499}.

\bibitem[{\citenamefont{Molla et~al.}(2017)\citenamefont{Molla, Chakrabarti,
  Debnath, and Mondal}}]{Molla-2016}
\bibinfo{author}{\bibfnamefont{A.~A.} \bibnamefont{Molla}},
  \bibinfo{author}{\bibfnamefont{S.~K.} \bibnamefont{Chakrabarti}},
  \bibinfo{author}{\bibfnamefont{D.}~\bibnamefont{Debnath}}, \bibnamefont{and}
  \bibinfo{author}{\bibfnamefont{S.}~\bibnamefont{Mondal}},
  \bibinfo{journal}{Astrophys. J.} \textbf{\bibinfo{volume}{834}},
  \bibinfo{pages}{88} (\bibinfo{year}{2017}), \eprint{1611.01266}.

\bibitem[{\citenamefont{Steiner et~al.}(2011)\citenamefont{Steiner, Reis,
  McClintock, Narayan, Remillard, Orosz, Gou, Fabian, and
  Torres}}]{Steiner-2011}
\bibinfo{author}{\bibfnamefont{J.~F.} \bibnamefont{Steiner}},
  \bibinfo{author}{\bibfnamefont{R.~C.} \bibnamefont{Reis}},
  \bibinfo{author}{\bibfnamefont{J.~E.} \bibnamefont{McClintock}},
  \bibinfo{author}{\bibfnamefont{R.}~\bibnamefont{Narayan}},
  \bibinfo{author}{\bibfnamefont{R.~A.} \bibnamefont{Remillard}},
  \bibinfo{author}{\bibfnamefont{J.~A.} \bibnamefont{Orosz}},
  \bibinfo{author}{\bibfnamefont{L.}~\bibnamefont{Gou}},
  \bibinfo{author}{\bibfnamefont{A.~C.} \bibnamefont{Fabian}},
  \bibnamefont{and} \bibinfo{author}{\bibfnamefont{M.~A.~P.}
  \bibnamefont{Torres}}, \bibinfo{journal}{Mon. Not. Roy. Astron. Soc.}
  \textbf{\bibinfo{volume}{416}}, \bibinfo{pages}{941} (\bibinfo{year}{2011}),
  \eprint{1010.1013}.

\bibitem[{\citenamefont{Steiner et~al.}(2012)\citenamefont{Steiner, McClintock,
  and Reid}}]{Steiner-2012}
\bibinfo{author}{\bibfnamefont{J.~F.} \bibnamefont{Steiner}},
  \bibinfo{author}{\bibfnamefont{J.~E.} \bibnamefont{McClintock}},
  \bibnamefont{and} \bibinfo{author}{\bibfnamefont{M.~J.} \bibnamefont{Reid}},
  \bibinfo{journal}{Astrophys. J. Lett.} \textbf{\bibinfo{volume}{745}},
  \bibinfo{pages}{L7} (\bibinfo{year}{2012}), \eprint{1111.2388}.

\bibitem[{\citenamefont{Sarkar and Chattopadhyay}(2018)}]{Sarkar-2018}
\bibinfo{author}{\bibfnamefont{S.}~\bibnamefont{Sarkar}} \bibnamefont{and}
  \bibinfo{author}{\bibfnamefont{I.}~\bibnamefont{Chattopadhyay}},
  \bibinfo{journal}{Int. J. Mod. Phys. D} \textbf{\bibinfo{volume}{28}},
  \bibinfo{pages}{1950037} (\bibinfo{year}{2018}), \eprint{1811.05947}.

\bibitem[{\citenamefont{Mitra et~al.}(2023)\citenamefont{Mitra, Ghoreyshi,
  Mosallanezhad, Abbassi, and Das}}]{Mitra-2023}
\bibinfo{author}{\bibfnamefont{S.}~\bibnamefont{Mitra}},
  \bibinfo{author}{\bibfnamefont{S.~M.} \bibnamefont{Ghoreyshi}},
  \bibinfo{author}{\bibfnamefont{A.}~\bibnamefont{Mosallanezhad}},
  \bibinfo{author}{\bibfnamefont{S.}~\bibnamefont{Abbassi}}, \bibnamefont{and}
  \bibinfo{author}{\bibfnamefont{S.}~\bibnamefont{Das}}, \bibinfo{journal}{Mon.
  Not. Roy. Astron. Soc.} \textbf{\bibinfo{volume}{523}}, \bibinfo{pages}{4431}
  (\bibinfo{year}{2023}), \eprint{2306.02453}.

\bibitem[{\citenamefont{Singh and Das}(2024)}]{Singh-2024-02256}
\bibinfo{author}{\bibfnamefont{M.}~\bibnamefont{Singh}} \bibnamefont{and}
  \bibinfo{author}{\bibfnamefont{S.}~\bibnamefont{Das}} (\bibinfo{year}{2024}),
  \eprint{2408.02256}.

\bibitem[{\citenamefont{Mitra et~al.}(2022)\citenamefont{Mitra, Maity,
  Dihingia, and Das}}]{Mitra-2022}
\bibinfo{author}{\bibfnamefont{S.}~\bibnamefont{Mitra}},
  \bibinfo{author}{\bibfnamefont{D.}~\bibnamefont{Maity}},
  \bibinfo{author}{\bibfnamefont{I.~K.} \bibnamefont{Dihingia}},
  \bibnamefont{and} \bibinfo{author}{\bibfnamefont{S.}~\bibnamefont{Das}},
  \bibinfo{journal}{Mon. Not. Roy. Astron. Soc.}
  \textbf{\bibinfo{volume}{516}}, \bibinfo{pages}{5092} (\bibinfo{year}{2022}),
  \eprint{2204.01412}.

\end{thebibliography}
%%%%%%%%%%%%%%%%%%%%%%%%%%%%%%%%%%%%%%%%%%%%%%%%%%%%%%%

\end{document}